\documentclass{elsarticle}
\usepackage{graphicx,color}
\usepackage{amsmath}
\usepackage{booktabs}

\usepackage{amsmath}
\usepackage{amssymb}

\def\nab{{\boldsymbol \nabla}}

\def\simle{\mathrel{\rlap{\raise 0.511ex \hbox{$<$}}{\lower 0.511ex \hbox{$\sim$}}}}
\def\simge{\mathrel{ \rlap{\raise 0.511ex \hbox{$>$}}{\lower 0.511ex \hbox{$\sim$}}}}
 
\def\q{{\boldsymbol q}}

\newcommand{\rmd}{{\rm d}}

\newcommand{\nn}{\nonumber\\ }

\newcommand \beq{\begin{eqnarray}}
\newcommand \eeq{\end{eqnarray}} 
\def\simle{\mathrel{\rlap{\raise 0.511ex \hbox{$<$}}{\lower 0.511ex 
\hbox{$\sim$}}}}
\def\simge{\mathrel{ \rlap{\raise 0.511ex 
\hbox{$>$}}{\lower 0.511ex \hbox{$\sim$}}}}
\newcommand{\del}{\partial}

\def\x{{\boldsymbol x}}

\def\q{{\boldsymbol q}}
\def\p{{\boldsymbol p}}

\begin{document}
\begin{frontmatter}
\title{Gluon Transport Equation with Effective Mass and Dynamical Onset of Bose-Einstein Condensation}

\author[cea]{Jean-Paul Blaizot}
\author[ind]{Yin Jiang}
\author[ind,rbrc]{Jinfeng Liao}

\address[cea]{Institut de Physique Th\'{e}orique (IPhT),  CNRS/URA2306, CEA Saclay,\\
F-91191 Gif-sur-Yvette, France}
\address[ind]{Physics Department and Center for Exploration of Energy and Matter,
Indiana University, 2401 N Milo B. Sampson Lane, Bloomington, IN 47408, USA}
\address[rbrc]{RIKEN BNL Research Center, Bldg. 510A, Brookhaven National Laboratory,
   Upton, NY 11973, USA}

\begin{abstract}
We study the transport equation describing a dense system of gluons, in the small scattering angle approximation, taking into account medium-generated effective masses of the gluons. We focus on the case of overpopulated systems that are driven to Bose-Einstein condensation on their way to thermalization. The presence of a mass modifies the dispersion relation of the gluon, as compared to the massless case,  but it is shown that this does not change qualitatively the scaling behavior in the vicinity of the onset.   
\end{abstract}
 \end{frontmatter}

\section{Introduction}

In previous papers \cite{Blaizot:2011xf,Blaizot:2013lga} it was argued that a dense system of gluons such as those created in the early stages of an ultra-relativistic heavy ion collision, could be driven to Bose-Einstein condensation, as the system evolves towards thermal equilibrium. This was inferred from a detailed study of the kinetic equation that takes into account 2 to 2 scattering, in the small scattering angle approximation. Overpopulation means that the dimensionless number $n/\epsilon^{3/4}$, where $n$ is the number density and $\epsilon$ the energy density, exceeds its value in equilibrium. An overpopulated system has too many gluons, relative to its total energy, to be accommodated in a Bose-Einstein distribution, and thermal equilibrium requires the formation of a condensate. 

Of course, such a condensate will develop provided the approach to thermal equilibrium proceeds with conservation of both energy and particle number. While energy is certainly conserved, inelastic processes of various kinds may change the  number of gluons (see e.g. \cite{Mueller:2005un,Blaizot:2014jna})\footnote{Note that quark production, although it decreases the number of gluons, does not necessarily hinds the formation of a condensate\cite{Blaizot:2014jna}.}, eventually preventing  the formation of a condensate in the true equilibrium state. However, particle number may be approximately conserved during much of the evolution, and this could be enough to approach condensation. Indeed, transport calculations indicate that the amplification of soft modes is a very rapid process, that a chemical potential is indeed dynamically generated and that the onset for condensation can be reached on short time scales. This is confirmed by calculations using the small angle approximation \cite{Blaizot:2013lga}, as well as complete solution of the Boltzmann equation \cite{Scardina:2014gxa,Xu:2014ega}. There are also indications that inelastic processes could accelerate the amplification of soft modes \cite{Huang:2013lia}, while the authors of Refs.~\cite{Kurkela:2012hp} and \cite{York:2014wja} seemingly reach a different conclusion. 

Clearly the analysis of inelastic scatterings requires further work, but this is beyond the scope of this paper. Motivation for studying the possibility for gluons to condense comes of course from the desire to better understand how matter produced in a high energy nucleus-nucleus collision evolves towards local thermal equilibrium (see \cite{Berges:2012ks,Huang:2014iwa} for a recent review). But, as we already emphasized in \cite{Blaizot:2013lga}, the general issue of the dynamical formation of a condensate  is an interesting problem in itself. It is of relevance in the context of cosmology (see e.g. \cite{Semikoz:1994zp}), or cold atom physics (see e.g. \cite{atomic_bec}). It has be studied using kinetic theory, or classical field simulations (see e.g. \cite{Berges:2013eia,Gelis:2013rba}). In the context of Quantum Chromodynamics,  the nature of the condensate  remains an interesting, but unsolved, question (for a recent study in a related context, see  \cite{Gasenzer:2013era}).

Our goal in this paper is to pursue our general study of the phenomenon within kinetic theory, using a transport equation that incorporates properly the effects of Bose statistics. The fact that the interactions are long range interactions validate the use of the  small angle approximation which reduces the transport equation to a Fokker-Planck equation, much easier to solve than the Boltzmann equation, thereby providing more direct analytical insight. This paper, as well as a companion paper, addresses issues that were not discussed in Ref.~\cite{Blaizot:2013lga} which is limited to the study of the onset of condensation. 
We want to extend our work so as to be able to obtain a complete  dynamical description of the approach to equilibrium including the formation of a condensate. In order to do so, we need to attribute  finite masses to the gluons. Such masses are automatically generated by the coupling to thermal fluctuations, and the proper transport equations that incorporate such self-energy corrections could be derived from first principles. However, for the purpose of the present study, it is sufficient to just give the gluons a mass, and correct appropriately the scattering matrix element. In fact two masses will be introduced. The screening mass $m_D$ regulates the infrared behavior of the collision kernel. The other mass, $m$,  modifies the dispersion relation, and one of the issues that we want to study is how this modification changes the onset of Bose-Einstein condensation. We shall see that, in fact, qualitatively  it does not. Finally, the role of the mass $m$ is to allow a clear definition of the equations that describe the evolution of the system beyond the onset, that is, in the presence of the condensate. This will be discussed in a companion paper \cite{BLM2}.

The outline of this paper is as follows. In the next section, we re-derive the approximate form of the transport equation in the small scattering angle approximation, paying attention to the presence of finite masses, and emphasizing the differences with the massless case discussed in \cite{Blaizot:2013lga}. In section 3, we present results of numerical solutions of the transport equation, illustrating the role of the mass on generic features of thermalization in both the underpopulated and overpopulated situations. In section 4, we focus on the critical regime that accompanies the onset of condensation. We show in particular that the change of the dispersion relation, from ultra-relativistic in the massless case, to non-relativistic in the massive case, does not change qualitatively the scaling regime. Details on the analytic calculation, as well as on the numerical solution, are given in two appendices.

\section{Derivation of the Transport Equation in the Massive Case}

In this Section, we derive the transport equation, under the  approximation of small scattering angle, taking into account  medium-induced effective masses for both the colliding  gluons and the exchange gluons.

\subsection{Generalities}

As in our previous papers \cite{Blaizot:2011xf,Blaizot:2013lga}, we assume  a transport equation for the single particle distribution of the following form
\begin{eqnarray} \label{Eq_transport}
  {\mathcal D}_t f_1&  = & C[f] \nonumber \\
  & =  &   \frac{1}{2} \int {{d^3\p_2} \over {(2\pi)^3 2 E_2}}{{d^3\p_3} \over {(2\pi)^3 2 E_3}}
   {{d^3\p_4} \over {(2\pi)^3 2 E_4}} {1 \over {2E_1}}  \mid {\cal M}_{12\to 34} \mid^2  \nonumber \\
   & \times & (2\pi)^4 \delta(P_1 + P_2 - P_3 -P_4) \nonumber \\
   &\times &  \{f_3f_4(1+f_1)(1+f_2) -  f_1f_2(1+f_3)(1+f_4)  \} 
  \end{eqnarray}
    where
    \begin{equation}
    {\mathcal D}_t \equiv \partial_t + {\boldsymbol v}_1 \cdot \nab ,   
      \end{equation} 
    and the factor $1/2$ in front of the integral is a symmetry factor. Summation over color and polarization is performed on the gluons 2,3,4; average over color and polarization is performed for gluon 1. The distribution function $f$ is a scalar object (i.e., independent of color and spin):
    \beq
    f(\x,\p)=\frac{(2\pi)^3}{2(N_c^2-1)}\frac{dN}{d^3\x d^3\p},
    \eeq
where $N$ denotes the total number of gluons. In other words, $f$ denotes the number of gluons of a given spin and color in the phase-space element $d^3\x d^3\p/(2\pi)^3$. We consider a uniform system, so that $f$ is in fact independent of $\x$. Also, in this paper, we consider a non-expanding system, so that ${\mathcal D}_t=\partial_t$. Finally, $P_i$ denotes the four-momentum of particle $i$, $P_i=(E_{\p_i},\p_i)$.

In this paper, contrary to our previous work, we assume that the gluons carry a small mass, arising from their interactions with the medium. This is a crude approximation to the true self-energy, but our main goal here is not to give a quantitative description of the phenomenon, but rather to study how the onset of Bose-Einstein condensation is affected by such a mass. We believe that a more complete, but much more difficult, treatment would not change our main conclusions. The major change brought by the presence of the mass is that of the dispersion relation of the modes. Thus the energy of a particle with momentum $\p$ is   $E_\p=\sqrt{\p^2+m^2}$. 
     
It is convenient to express the final momenta in terms of the initial ones and of the momentum $\q$ transferred in the collision. We set 
\beq\label{transfernot}
\p_3=\p_1+\q,\qquad \p_4=\p_2-\q.
\eeq
One can then perform the integrations over $\p_3$ and $\p_4$ using the   three-momentum delta function. We get
\beq
C[f]=  \int_{\p_2} \int_{\q}   w(\p_1,\p_2,\q)\{f_{12\to 34}\},
 \eeq
with
\beq
w(\p_1,\p_2,\q)\equiv \frac{\pi \delta(E_1+E_2-E_3-E_4)}{16E_1 E_2 E_3 E_4}\mid {\cal M}_{12\to 34} \mid^2  ,
\eeq  
and 
\beq
\{f_{12\to 34}\}\equiv  f_3f_4(1+f_1)(1+f_2) -  f_1f_2(1+f_3)(1+f_4),
\eeq
where $f_i$ is a shorthand notation for $f_{\p_i}$. The quantity $w(\p_1,\p_2,\q)$ may be interpreted as the rate of collisions of particle 1 with particle 2, in which momentum $\q$ is transferred to particle 1, its momentum $\p_1$ becoming $\p_1+\q$. Note that the symmetry factor is included in the definition of $w$. 
The (dimensionless) matrix element ${\cal M}_{12\to 34}$ will be discussed shortly. 
It is understood in the expressions above that the momenta $\p_3$ and $\p_4 $ are expressed in terms of $\p_1,\p_2,\q$ according to (\ref{transfernot}). Also, we used the shorthand for momentum integration
\beq
\int_{\p}\equiv \int\frac{\rmd^3 \p}{(2\pi)^3}.
\eeq
Note that the symmetries of the matrix element (see below) entail the property
\beq\label{symmetriesofME}
w(\p_1,\p_2,-\q)=w(\p_1,\p_2,\q)=w(\p_2,\p_1,-\q).
\eeq

\subsection{The small scattering angle approximation and the Fokker-Planck equation}

Under the small scattering angle approximation, the momenta of incident particles get changed very little during each collision, and the kinetic equation can be approximated by a Fokker-Planck equation in momentum space \cite{Mueller:1999fp}. Following a standard procedure \cite{LifshitzPitaevskii}, we write the collision integral as
\beq
C[f]=-\nab\cdot {\cal J}=-\frac{\del {\cal J}_i}{\del p_i}.
\eeq
Note the gradient $\nab$ in the above, and for the rest of this paper, is the momentum space gradient i.e. $\nab=\nab_{\p}$.  
In order to estimate ${\cal J}_i$, the component of the current (of particles 1) in the direction $i$ (with $i=1,2,3$),  we count the number of particles that, as a result of collisions during the interval $\rmd t$, cross  a surface element orthogonal to the direction $i$ and located at a particular value $\bar\p$  of $\p_1$.  An elementary analysis yields
\begin{eqnarray}
{\cal J}_{{i}} &=& \int_{\q,\,q_i>0} \int_{\p_2}\, \int_{\bar p_i-q_i}^{\bar p_i}dp_i \, w(\p_1,\p_2,\q)\,\nn
 &\times&  {\bigg \{ }  f_{\p_1}\, f_{\p_2}\, (1+f_{\p_1+\q})\, (1+f_{\p_2-\q})   - (1+f_{\p_1})\, (1+f_{\p_2})\, f_{\p_1+\q}\, f_{\p_2-\q} {\bigg \}},\nn
\end{eqnarray}
where the $\q$-integration is restricted to positive components $q_i$. 
In the small angle approximation the combination of statistical factors simplifies into
\begin{eqnarray}
&& f_{\p_1}\, f_{\p_2}\, (1+f_{\p_1+\q})\, (1+f_{\p_2-\q})   - (1+f_{\p_1})\, (1+f_{\p_2})\, f_{\p_1+\q}\, f_{\p_2-\q}  \nonumber \\
&& \approx \q \cdot \left[ h_{\p_1}(\nab f)_{\p_2} - h_{\p_2}(\nab f)_{\p_1}  \right] + {\cal O}(\q^2),
\end{eqnarray}
where we have introduced the notation $h_{\p} \equiv f_{\p}(1+f_{\p})$. 
In addition, we notice that, in this approximation, the energy conservation  implies
\begin{eqnarray} \label{Eq_en_cons}
0 =  \q \cdot {\boldsymbol v}_1 - \q \cdot {\boldsymbol v}_2  + {\cal O}(\q^2).
\end{eqnarray}
By taking all these together we obtain the following leading order expression for the momentum flux:
\begin{eqnarray} \label{Eq_flux}
{\cal J}^i= \frac{1}{2} \int_\q \int_{\p_2}\, \q^i \,
w(\p_1,\p_2\, ,\q)\, \q\, \cdot \left[ h_{\p_1}(\nab f)_{\p_2} - h_{\p_2}(\nab f)_{\p_1}  \right].
\end{eqnarray}
Note that in the above equation we have relaxed the constraint $q_i>0$ on the $\q$-integration, dividing  the result by $2$ (using the fact that the integrand is even in $\q$, see Eqs.~(\ref{symmetriesofME})).  We can rewrite the current as follows 
\beq\label{currentJi}
{\cal J}^i= \int_{\p_2}\, B^{ij}(\p_1,\p_2)
 \left[ h_{\p_1}(\nab^j f)_{\p_2} - h_{\p_2}(\nab^j f)_{\p_1}  \right],
\end{eqnarray}
with the (dimensionless) angular tensor
\beq
B^{ij}(\p_1,\p_2)=\frac{1}{2}\int_\q \q^i\q^j w(\p_1,\p_2,\q).
\eeq

\subsection{Conservation laws}

The particle number conservation is obvious due to the structure of the collision term as the divergence of a current:
\begin{eqnarray}
\int_{\p_1} C[f_{\p_1}] = 0
\end{eqnarray}
To prove the energy conservation requires a little more effort
\begin{eqnarray}
\int_{\p_1} E_{\p_1} C[f_{\p_1}] &=&  - \int_{\p_1} E_{\p_1} \nab_{\p_1} \cdot {\cal J}(\p_1) \nonumber \\
&=& - \int_{\p_1}  \nab_{\p_1} \cdot \left[E_{\p_1}\, {\cal J}(\p_1)\right] + \int_{\p_1} \left(\nab_{\p_1} E_{\p_1}\right) \cdot {\cal J}(\p_1) \quad \nonumber \\
&=& \int {{d^3 \p_1} \over {(2\pi)^3}}\, {\boldsymbol v}_1 \cdot {\cal J}(\p_1) \nonumber \\
&=& \frac{1}{2} \int_{\p_1} \int_{\p_2}  \int_\q\, w(\p_1,\p_2\, ;\q) \nonumber \\
&& \qquad  \times ({\boldsymbol v}_1 \cdot \q) \,\, \q\, \cdot \left[ h_{\p_1}(\nab f)_{\p_2} - h_{\p_2}(\nab f)_{\p_1}  \right] \quad \nonumber \\
&=& \frac{1}{2} \int_{\p_1} \int_{\p_2}  \int_{\q}\, w(\p_1,\p_2\, ;\q) \nonumber \\
&& \qquad  \times \left[({\boldsymbol v}_1 \cdot \q) \, h_{\p_1}( \q\, \cdot \nab f)_{\p_2} - ({\boldsymbol v}_2 \cdot \q) \, h_{\p_2}( \q\, \cdot\nab f)_{\p_1}  \right], \quad \nonumber \\
&=& 0
\end{eqnarray}
where in the last steps we have used  Eq.(\ref{Eq_en_cons}) and Eqs.~(\ref{symmetriesofME}).

\subsection{The matrix element}

 The matrix element for (in vacuum) gluon-gluon scattering ($1+2\to 3+4$) reads (spin and color averaged for 1, and summed for 2,3,4)
 \begin{eqnarray}
 \mid {\cal M} \mid^2 = 128\pi^2\alpha_s^2 N_c^2 \left[ 3- \frac{t\, u}{s^2} -  \frac{s\, u}{t^2} - \frac{t\, s}{u^2} \right],\qquad  128\pi^2\alpha_s^2 N_c^2=72 g^4,
 \end{eqnarray}
 with $s,t,u$ the standard Mandelstam variables:
 \beq s=(P_1+P_2)^2,\qquad t=(P_1-P_3)^2,\qquad u=(P_1-P_4)^2.
 \eeq

In the small scattering angle approximation, the dominant contributions come from the kinematic regions $t\approx 0$ and $u\approx 0$. With the vacuum matrix element and in massless case, one would have the following approximation: 
\beq
\mid {\cal M} \mid^2 = 72 g^4 \left[ 3- \frac{t\, u}{s^2} -  \frac{s\, u}{t^2} - \frac{t\, s}{u^2} \right]\approx 72 g^4 \left[ -  \frac{s\, u}{t^2} - \frac{t\, s}{u^2} \right]\approx 144 g^4\frac{s^2}{t^2},
\eeq
where in the last step we have used the fact that the two contributions are equal, and, for massless particles,  $u=-(s+t)\approx -s$ (for $ t\approx 0$).

Now we consider the modifications of the matrix elements that need to be taken into account when the gluons are massive. There are two distinct physical effects. One is the screening of the $t$-channel singularity. This should be taken care of by  separating the transverse and the longitudinal channels, and including the proper polarization tensors in the exchange gluons. In this paper we simply modify the denominator by substituting $t\to t-m_D^2$, with $m_D$ a screening mass whose main role here is that of a regulator. The other physical effect is coming from self-energy corrections on the external lines. As already mentioned, we simply take these into account by giving the gluon a small thermal mass $m$,  allowing $m$ to differ from $m_D$. In summary, we replace the matrix element derived for massless particles by
 \beq
\mid {\cal M} \mid^2  \to  144 g^4 \frac{s ^2}{(t-m_D^2)^2},\qquad s\approx 2E_1E_2(1-{\boldsymbol v}_1\cdot{\boldsymbol v}_2),
\eeq
 where ${\boldsymbol v}_i=\p_i/E_i$ is the velocity of particle $i$, and since we are interested in the small $m$ limit, we dropped a  term $\sim m^2$ in the expression of $s$. With that, we then obtain, in the leading order of the small scattering angle approximation 
\begin{eqnarray}
w =36 \pi {g^4}\delta (\q \cdot {\boldsymbol v}_1 - \q \cdot {\boldsymbol v}_2)\frac{(1-{\boldsymbol v}_1\cdot{\boldsymbol v}_2)^2}{(\omega^2-\q^2 - m_D^2)^2}
\end{eqnarray}
where $\omega={\boldsymbol v}_1\cdot\q$.

With this matrix element, the angular tensor takes the form
\beq\label{angulartensor}
B^{ij}(\p_1,\p_2) =18\pi g^4 \int_{\q} \,\delta({\boldsymbol v}_1\cdot\q-{\boldsymbol v}_2\cdot\q)\q^i\q^j \frac{(1-{\boldsymbol v}_1\cdot{\boldsymbol v}_2)^2}{(\omega^2-\q^2 - m_D^2)^2}.
\eeq

\subsection{The isotropic case}

We assume that  the distribution function is a function of the energy, $f_{\p} = f(E_\p)$ with $E_\p=\sqrt{\p^2+m^2}$. In such case we have the following relation
\begin{eqnarray}
\nab f_{\p} = {\boldsymbol v} \, {f}'(E_\p),\qquad {f}'(E) \equiv \frac{\rmd f(E)}{ \rmd E},
\end{eqnarray}
with the velocity  given by ${\boldsymbol v} = \p /E_\p = \nab_{\p} E_\p $. 
 Then we have
\beq
  h_{\p_1}(\nab^j f)_{\p_2} - h_{\p_2}(\nab^j f)_{\p_1}  ={\boldsymbol v}_2^j \, h_1  f^\prime_2-{\boldsymbol v}_1^j h_2 f^\prime_1 ,
 \eeq
 where we introduced the simplified notation $f_i=f(E_i)=f_{\p_i}$ (and similarly for $h_i$) that will  be used throughout the paper. 

In this isotropic case, the calculation of the angular tensor (\ref{angulartensor}) simplifies. In particular,  the current $\vec{\cal J}(\p)$ is a vector aligned with the direction of $\p$, that is (with $p_1=|\p_1|$ and $\hat\p_1=\p_1/p_1$)
\beq
\vec{\cal J}(\p_1)=\hat\p_1{\cal J}(p_1). 
\eeq
It follows that the kinetic equation can be written as 
\beq\label{spehricaleqn}
{\mathcal D}_t f_1= - \frac{1}{p_1^2}  \partial_{p_1} 
 \left\{ p_1^2 {\cal J}(p_1)\right\}.
\eeq
The  calculation presented in  Appendix A yields 
\beq\label{smallpJ}
 {\cal J}(p_1)=36 \pi\alpha_s^2 \int_{\p_2}\,({h_1}{f^\prime_2} - {h_2}{f^\prime_1}) {\cal Z}(v_1,v_2),
\eeq
where the explicit expressions of the dimensionless function ${\cal Z}(v_1,v_2)$ is given in Appendix A.

\subsection{The massless limit}

It is shown  in~\ref{Sect:smallmomentumJ} that  the current  at small momentum has the following structure
\beq\label{currentmassive}
 -{\cal J}(p_1 \to 0)&\simeq &   36\pi \alpha_s^2 {\cal L}  
 \left[{I}_a(p_1)  \partial_{p_1}{f}_1 + \frac{p_1}{E_1}\, {I}_b(p_1) h_1 \right] , 
\end{eqnarray}
 where $\cal L$ is a positive constant, and  (see Eq.~(\ref{Iap1Ibp1}))
 \beq\label{integralstilde}
 {I}_a(p_1) \equiv \frac{p_1}{m} \int_{\p_2} \frac{h_2}{v_2}, \quad    { I}_b(p_1) \equiv - \frac{p_1}{m} \int_{\p_2} \frac{{f}^\prime_2}{v_2}  = - \frac{p_1}{m} \int_{\p_2} \,\frac{1}{v_2^2} \partial_{p_2}{f}_2.
 \eeq
This structure is identical to that obtained in the massless limit \cite{Blaizot:2013lga}, with the integrals $I_a$ and $I_b$ given by
\beq\label{IaIb}
I_a=\int_\p f(p)(1+f(p),\qquad I_b=\int_\p \frac{2f(p)}{p}.
\eeq

 We shall end this section by making a general comment on the Fokker-Planck equation in this massless limit, in order to clarify the interpretation of the two competing terms that are present in the current. For simplicity we  focus on the isotropic case so that the kinetic equation is of the form\footnote{We absorb here the constant factor $36\pi \alpha_s^2 {\cal L}$ into the redefinition of time (see Eq.~(\ref{timescale}) below).}
\begin{eqnarray}
\partial_t f(p) = \frac{1}{p^2} \partial_p \left\{ p^2 \left[ I_a \partial_p f(p) + I_b f (1+f)  \right] \right\}.
\end{eqnarray}
Note that energy conservation, $\partial_t\int_\p pf(p)=0$, entails
\beq
I_a\int_\p \frac{\partial f}{\partial p}=I_b\int_\p f(1+f), 
\eeq
a relation that is obviously satisfied given the definitions (\ref{IaIb}) of  $I_a$ and $I_b$. 

To gain insight into the physical meaning of the two terms on the r.h.s. of this equation, let us multiply  both sides by $p^2$ and integrate over momentum. We obtain
\begin{eqnarray} 
\partial_t \int_{\p} p^2 f(p) = \int_{\p} p^2 \frac{1}{p^2} \partial_p \left\{ p^2 \left[ I_a \partial_p f(p) + I_b f (1+f)  \right] \right\}.
\end{eqnarray}
Using integration by part for the r.h.s., we  eventually obtain 
\begin{eqnarray} \label{eq_diffusion_drag}
\partial_t \langle p^2\rangle = 6I_a \langle n\rangle - 2 I_b \tilde{ \langle p\rangle},
\end{eqnarray} 
where 
\begin{eqnarray}
 \langle p^2\rangle \equiv   \int_{\p} p^2 f(p), \qquad 
 \langle n\rangle  \equiv \int_{\p}  f(p) , \qquad
\tilde{ \langle p\rangle}  \equiv \int_{\p}  p f(1+f).
\end{eqnarray}
The equation (\ref{eq_diffusion_drag}) may be interpreted as follows. The first term on the r.h.s., $6I_a \langle n\rangle$, represents the diffusion of particles in momentum space that results from  multiple small-angle scatterings. If this first term was the only one, the  system would diffuse indefinitely, with $\langle p^2\rangle \sim I_a \langle n\rangle t $ at late time. The second term on the r.h.s., $- 2 I_b\tilde{\langle p\rangle} $, represents a drag force, with the characteristic of a friction. It is negative as long as there is any nonzero momentum in the system and therefore  opposes the diffusive  contribution,  which is generically positive. In the absence of diffusion, this term would cause the   magnitude of the momentum to decrease continuously, eventually bringing all particles to a state of zero momentum. This drag term is the agent that drives the system towards condensation. 
Finally, with both terms present on the r.h.s of the equation,  thermal equilibrium can be reached, and   the system approaches the fixed point corresponding to a Bose-Einstein distribution, with $T=I_a/I_b$.

\section{Numerical solutions to the transport equation}

We turn now to the discussion of results obtained by solving numerically the Fokker-Plank equation for relevant cases. Details on the numerical procedure that we used can be found in~\ref{Sect:numerics}.  We shall write the transport equation in terms of dimensionless quantities. In order to do so, we express all momenta (and masses,  energies, temperature) in units of $Q_s$, and the time in units of $1/Q_s$. In fact, we also absorb into the dimensionless time a numerical factor, setting
\begin{eqnarray}\label{timescale}
t \equiv \left( \frac{1}{Q_s}\right)\frac{\tau\pi}{18\alpha_s^2}. 
\end{eqnarray}
This is the factor that should be used if one wants to relate the time $\tau$ of the simulation to the physical time $t$.~\footnote{Note that this factor differs by a factor $2\pi^2 {\cal L}$ from the ``natural'' factor that appears for instance in the expression (\ref{smallpJ}) of the current. Note that $\pi/(18\alpha_s^2)\approx 0.7$ so that $t\approx \tau/Q_s$.}

In the dimensionless variables, the transport equation reads \beq\label{spehricaleqn2}
{\mathcal D}_\tau f(p)= - \frac{1}{p^2}  \partial_{p} 
 \left\{ p^2 {\cal J}(p)\right\}, \qquad  {\cal J}(p_1)= \int_{\p_2}\,({h_1}{f^\prime_2} - {h_2}{f^\prime_1}) {\cal Z}(v_1,v_2).\nn
\eeq
Note that since from now on we shall be dealing mostly with dimensionless variables, we keep the same names for the dimensionful and dimensionless momenta. 
\begin{figure}[!htb]
\begin{center}
\includegraphics[width=6.5cm]{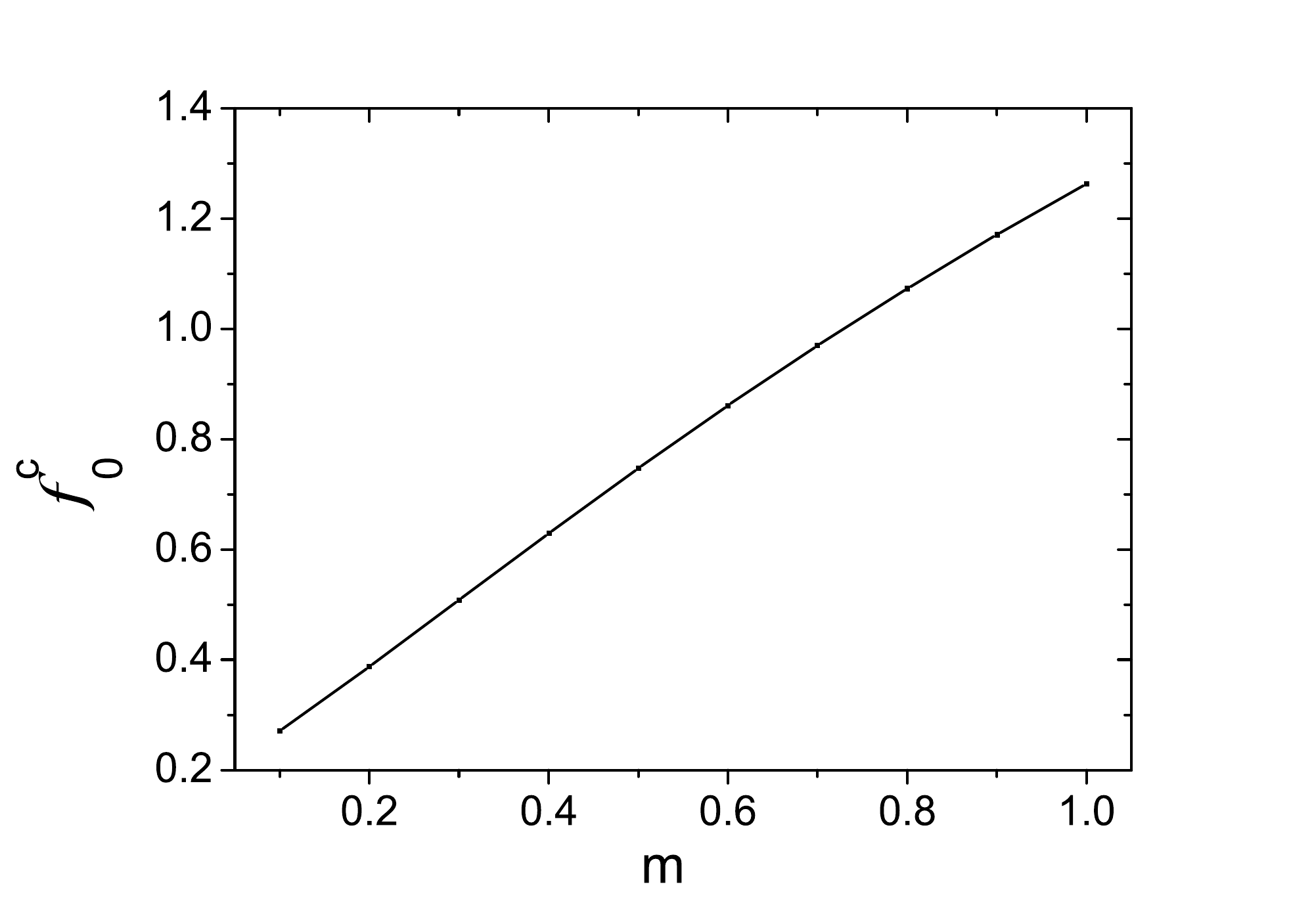}\includegraphics[width=6.5cm]{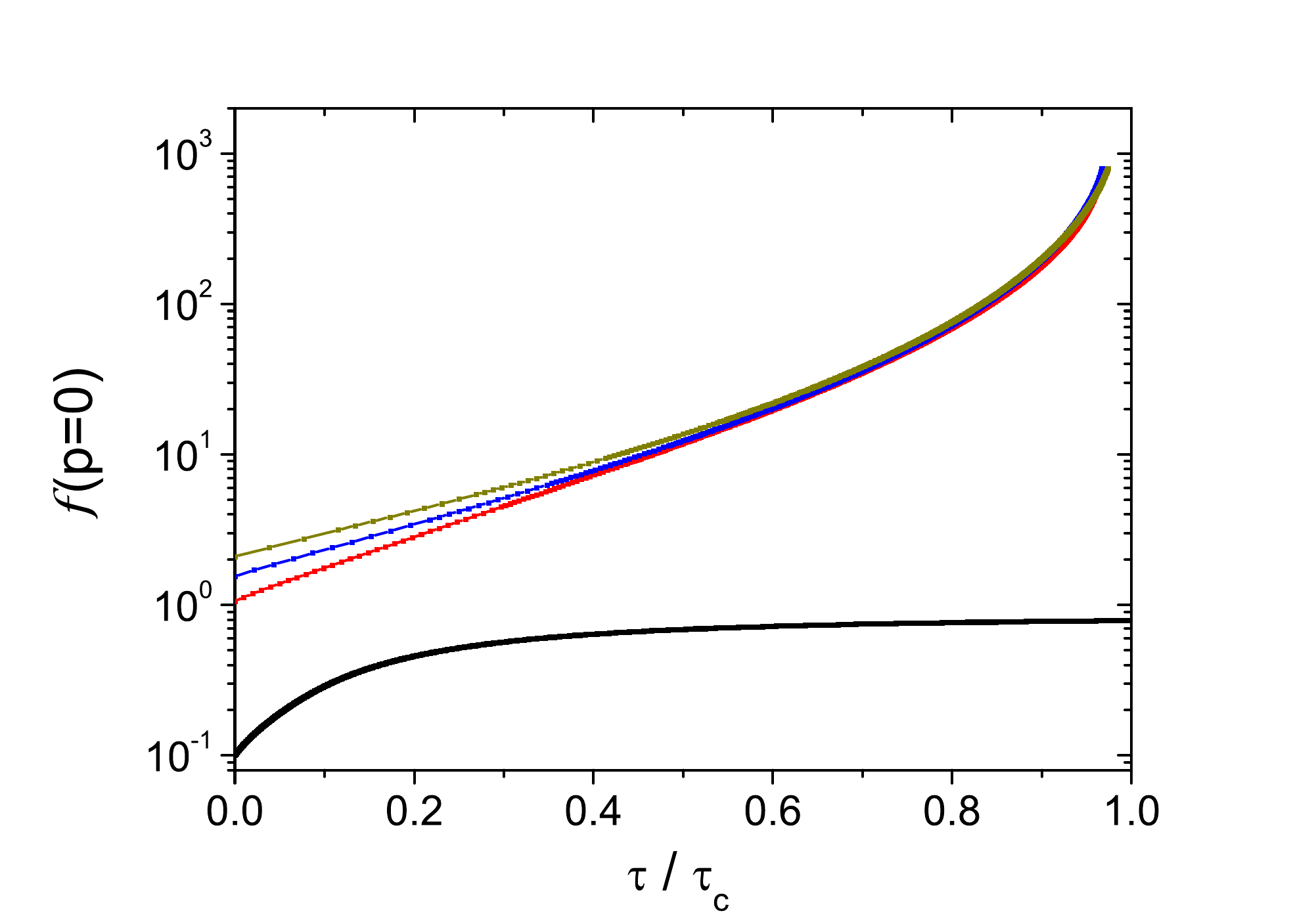}
\caption{(Color online.) Left: Critical $f_0$ as a function of the gluon mass $m$. Points above the curve correspond to overpopulated initial conditions, below the curve to underpopulated initial conditions. Right: The evolution of $f(p\approx 0)$ for different initial occupations $f_0$ (with $m=0.3$): the lower (black) curve is for underpopulated initial condition, and the other curves are for overpopulated initial conditions.}\label{foc}
\end{center}
\end{figure}

The initial condition is chosen be of the Glasma type \cite{Blaizot:2011xf}, i.e. $f(E)= f_0 (1+e^{10(E-1.5)})^{-1}$, with $E=\sqrt{{p}^2+m^2}$. For this family of initial conditions, there is a critical value of $f_0$, that we call $f_0^c$, above which the system becomes overpopulated. When $f_0=f_0^c$, the equilibrium distribution is a Bose-Einstein distribution with a maximal chemical potential $\mu=m$. The value of $f_0^c$ depends on the mass $m$ of the gluons, and its variation with $m$ is illustrated in Fig.~\ref{foc}. It is seen that $f_0^c$ increases with $m$: the value of $f_0$ required  to reach the onset of condensation is larger for massive particles than for massless ones. The right hand panel of Fig.~\ref{foc} illustrates the behavior of $f(p=0)$ in the two generic situations of underpopulation where $f(0)$ reaches a constant value and that of overpopulation where $f(0)$ diverges at  a finite time $\tau=\tau_c$.

Note that in all cases, the  infrared part of the distribution function evolves rapidly towards an approximate classical thermal distribution function, 
\beq
f(E) \to \frac{T^* }{ (E -\mu^*)}
\eeq
with an effective temperature $T^*$ and effective chemical potential $\mu^*$ that can be determined numerically. The evolution of $T^*$ and $\mu^*$ with time depends on whether the system is over or underpopulated, as we shall discuss in the following.
\begin{figure}[!htb]
\hspace{-0.5 cm}\includegraphics[width=6.5cm]{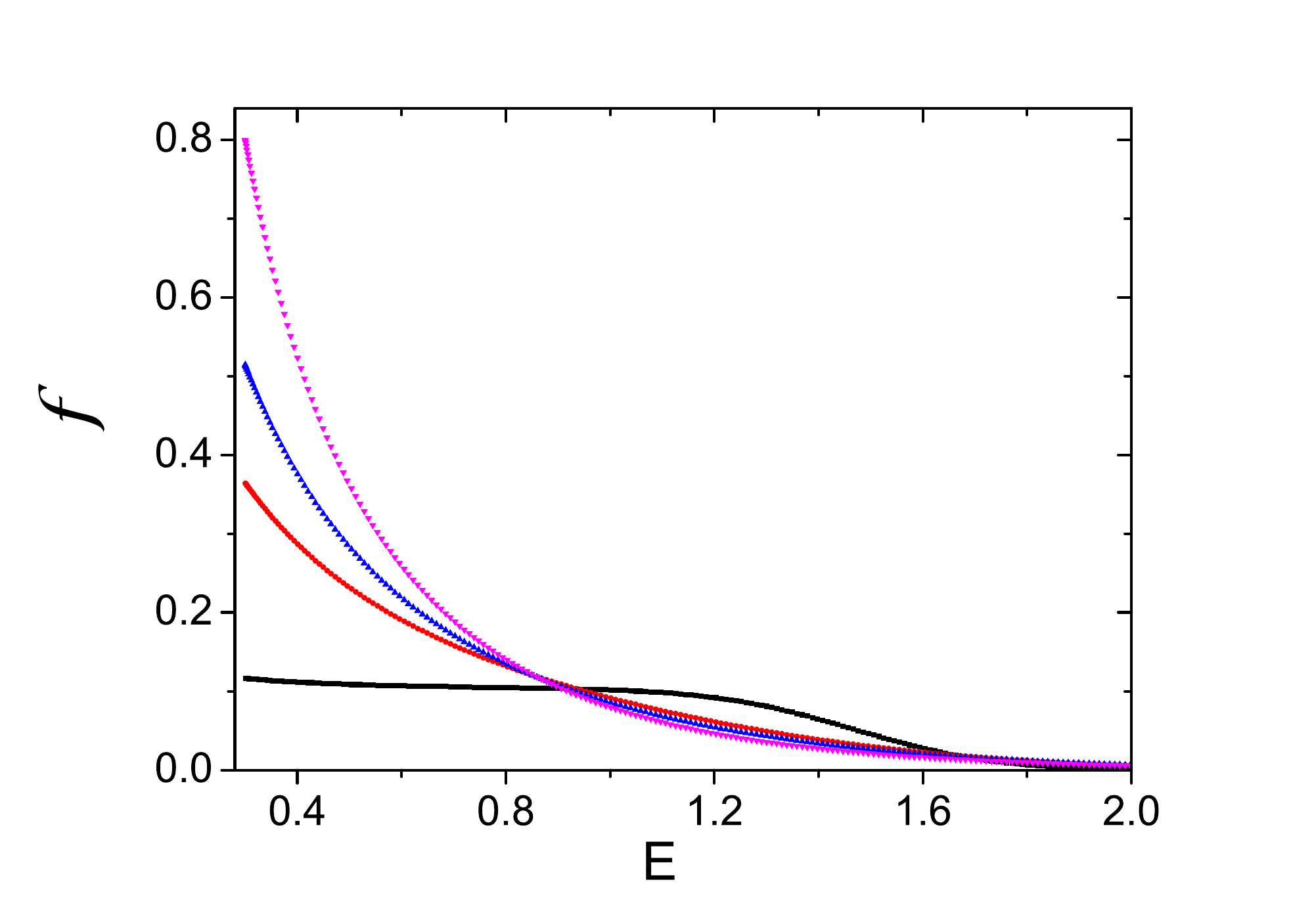}\includegraphics[width=6.5cm]{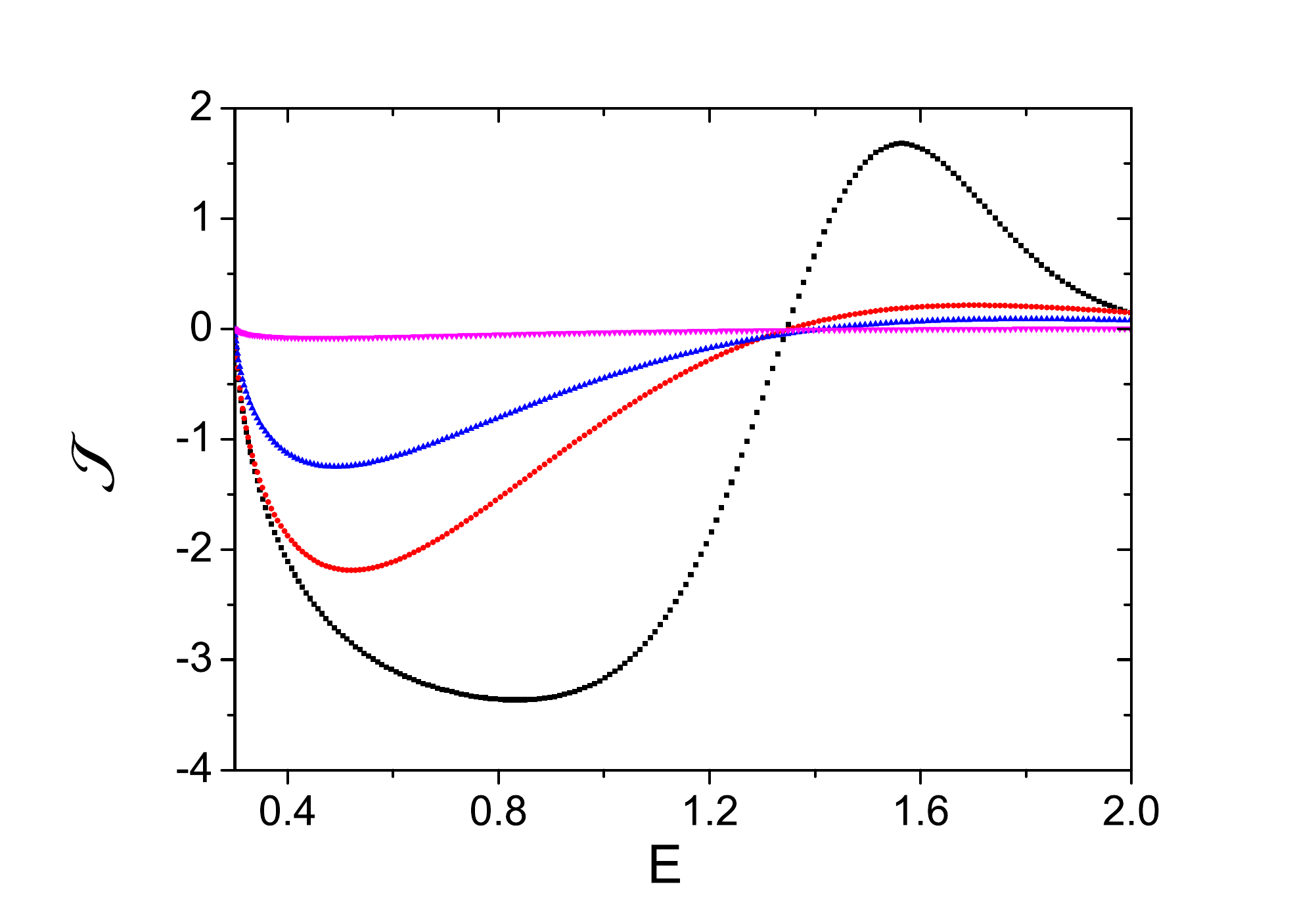}
\caption{Left: Distribution function $f(E)$ at various times $\tau=$0.05, 0.8, 1.4, 6.25 (from bottom to top at $E=0.4$). Right: Current ${\cal J}$  at $\tau=$0.05, 0.8, 1.4, 6.25 (from bottom to top at $E=0.4$). Underpopulated initial condition, with $m=0.3$ and $f_0=0.1$. }\label{fig2}
\end{figure}

\subsection{Underpopulated case}

\begin{figure}[!htb]
\begin{center}
\hspace{-0.6 cm}\includegraphics[width=6.5cm]{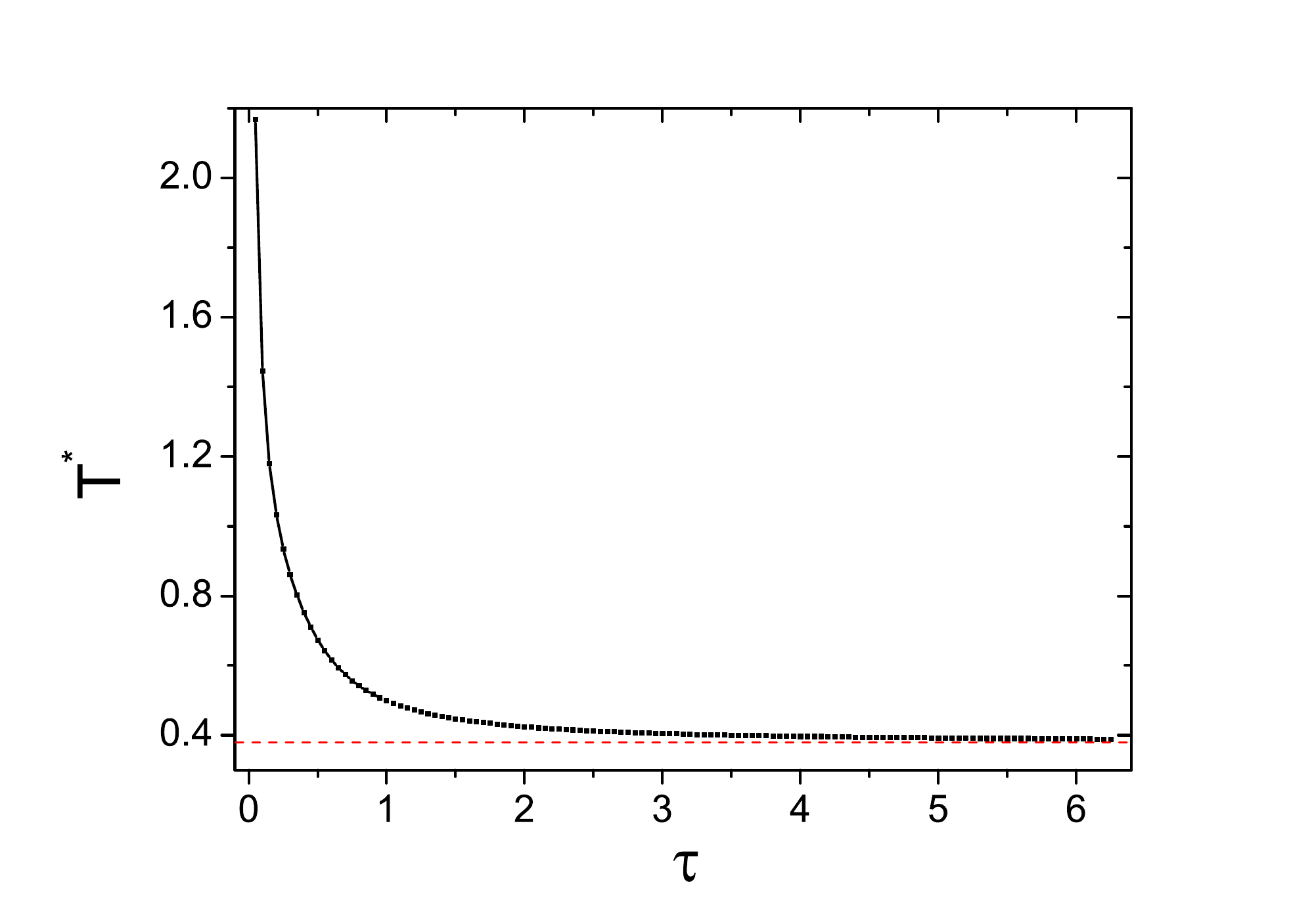}\includegraphics[width=6.5cm]{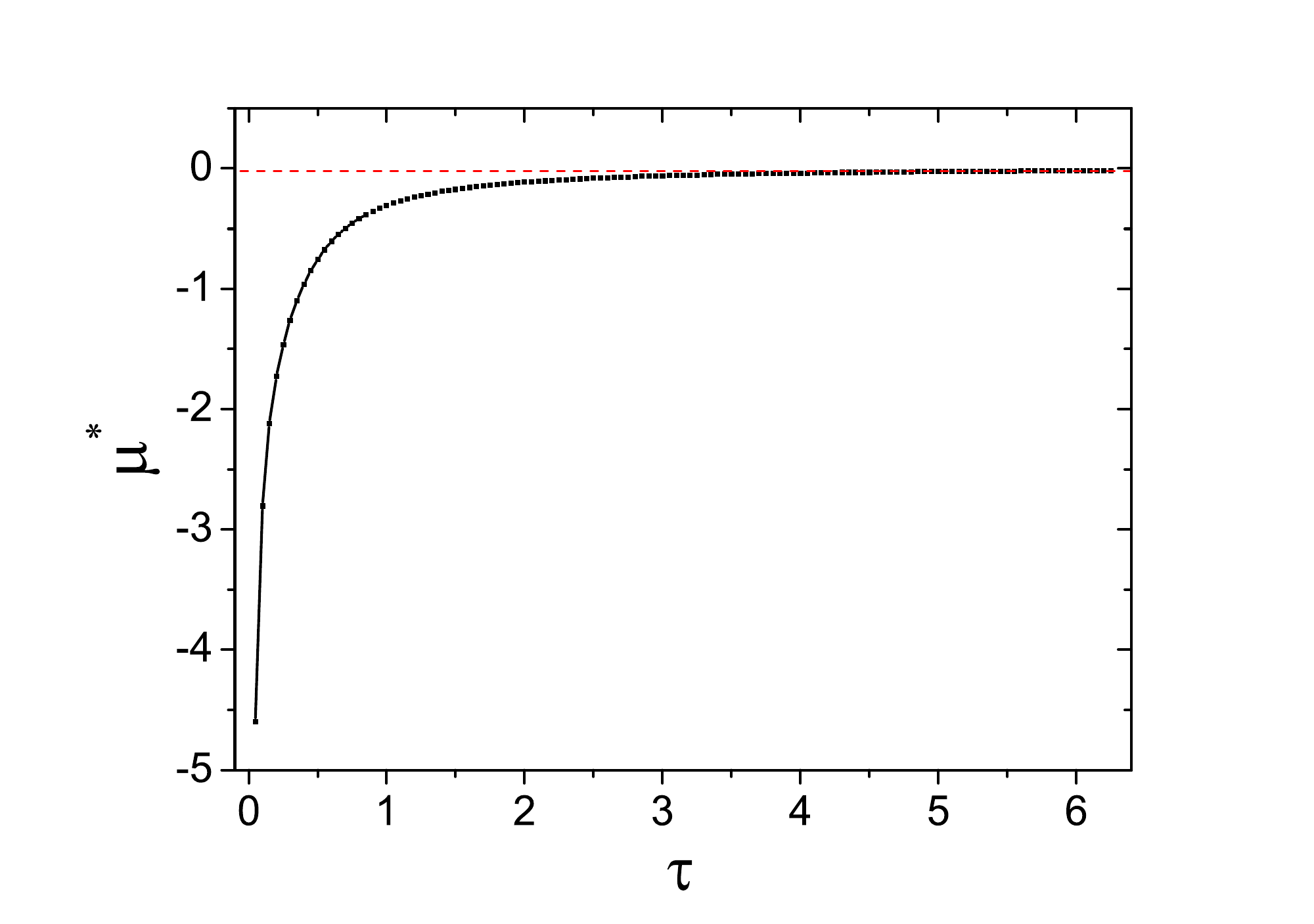}
\caption{(Color online.) The evolution  with time of the effective local temperature $T^*$ (left) and chemical potential $\mu^*$ (right) for an underpopulated initial condition ($m=0.3$ and $f_0=0.1$). At late time, both quantities approach their expected equilibrium values, indicated by the horizontal (red) dotted lines.}\label{fig3}
\end{center}
\end{figure}

Starting from the initial distribution corresponding to underpopulation, we verify that the system evolves to the expected equilibrium state.  
The  approach to  thermalization is illustrated in  Fig.\ref{fig2} which displays the evolution of the distribution function $f(p)$ and that of the current ${\cal J}(p)$. 
The current changes its sign around $E\simeq 1.3$, the point which separates the effects of the two competing components of the current. For larger values of $E$, the current is diffusive and positive. For smaller values of $E$, the current is dominated by its drag component, and is negative. As time progresses these two components of the current move particles in momentum space, the drag term pushing particles towards small momenta, the diffusion term smoothening the distribution at large momenta. The  resulting distribution gradually evolves towards the equilibrium distribution, as indicated in the left panel of Fig.~\ref{fig2}, while the current diminishes, and eventually vanishes (which takes place approximately for the largest time considered in the plot). The evolution with time of the effective parameters that characterize the infrared part of the distribution function are displayed in Fig.~\ref{fig3}. This figure clearly demonstrates that the system thermalizes as expected.

\subsection{Overpopulated cases}

\begin{figure}[!htb]
\begin{center}
\hspace{-0.6 cm}\includegraphics[width=9 cm]{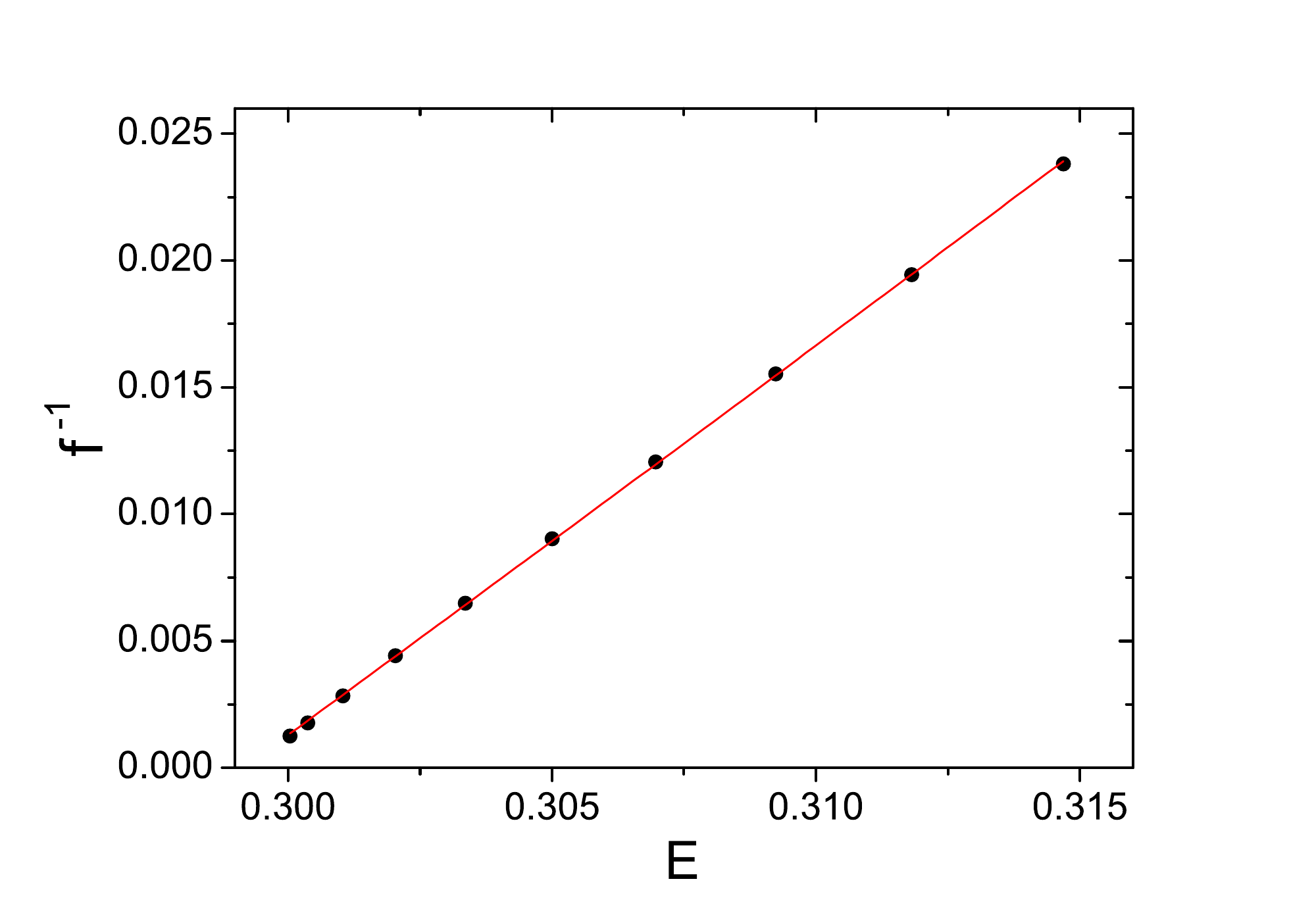}
\caption{The inverse of the distribution as a function of energy, $1/f$ versus $E$, for $m=0.3$, and $\tau=0.101$ corresponding to the onset of condensation. }\label{fig4}
\end{center}
\end{figure}

We turn now to the situation with overpopulation. We expect, from our previous work, that 
the system will approach the onset of condensation in a finite time $\tau=\tau_c$. We have studied the evolution   with different masses($m=0.1$, $0.3$, $0.5$, $0.7$). In the numerical solution, the time step is self-adaptive i.e. becoming smaller when the evolution becomes faster, which allows us to evolve the system in each case very close to the onset. The evolution is stopped when $f>800$ for the smallest momentum grid point, which roughly corresponds to the situation when the difference between the local chemical potential $\mu^*$ and the mass becomes less than the mesh size. That is, the evolution is stopped very near the onset of condensation where $\mu=m$.  An illustration of the energy dependence of the distribution near the onset is displayed in Fig.~\ref{fig4}

The Table~\ref{Table} gives some typical values of the parameters $f_0^c$ and $\tau_c$ for different initial conditions, and different values of the mass $m$ and the Debye mass $m_D$. One  notices that, for fixed masses (e.g. m=0.3)  the onset time decreases with increasing $f_0$, which is as expected: the larger the  overpopulation, the shorter time it takes to reach the onset of BEC. For fixed $m$ and $f_0$ an increase of the Debye mass leads to an increase of $\tau_c$ (compare for instance the lines corresponding to $m=0.7$ and $f_0=1$). This is because an increase of $m_D$ reduces the scattering rate, which slows down the evolution.  A larger increase of $\tau_c$ results from the increase of $m$ at fixed $f_0$ and $m_D$. The latter phenomenon is in line with what we observed earlier, namely that $f_0^c$ increases with $m$. 
\begin{table}[!htbp]
\begin{center}
\begin{tabular}{ccccc}
\toprule
m& $m_D$& $f_0$& $f_0^c$& $\tau_c$\\
\midrule
0.1& 0.1& 1.0& 0.271& 0.029\\
0.3& 0.3& 1.0& 0.508& 0.101\\
0.5& 0.5& 1.0& 0.747& 0.259\\
0.7& 0.7& 1.0& 0.969& 0.663\\
0.3& 0.3& 1.2& 0.508& 0.065\\
0.3& 0.3& 1.4& 0.508& 0.046\\
0.3& 0.3& 1.6& 0.508& 0.034\\
0.3& 0.3& 1.8& 0.508& 0.026\\
0.3& 0.3& 2.0& 0.508& 0.021\\
0.1& 0.7& 1.0& 0.271& 0.095\\
0.7& 0.1& 1.0& 0.969& 0.242\\
\bottomrule
\end{tabular}
\end{center}
\caption{Critical occupation factor $f_0^c$ and onset time scale $\tau_c$ for different combinations  of  parameters.}\label{Table}
\end{table}

\begin{figure}[!htb]
\begin{center}
\hspace{-0.6 cm}\includegraphics[width=6.5 cm]{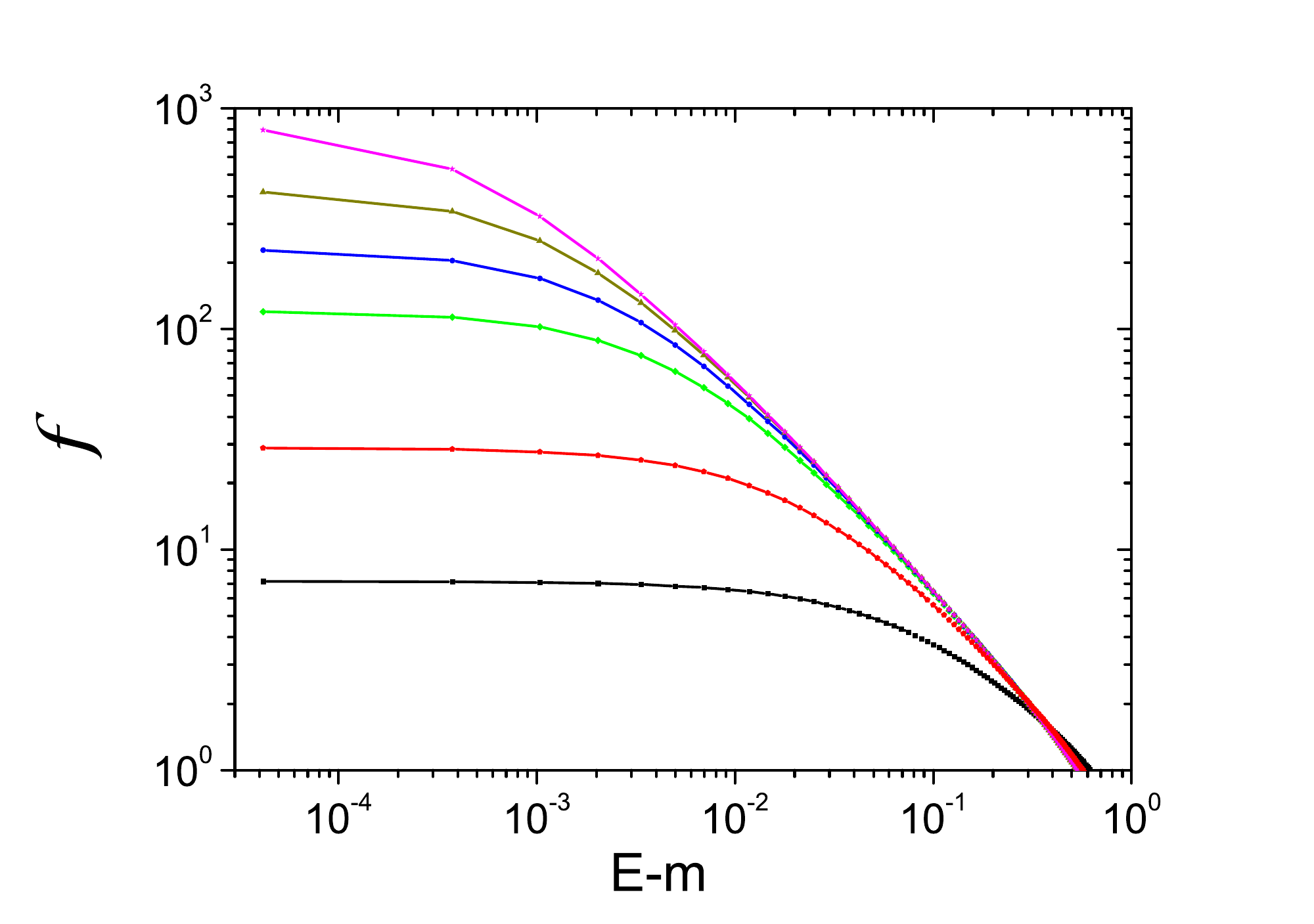}\includegraphics[width=6.5cm]{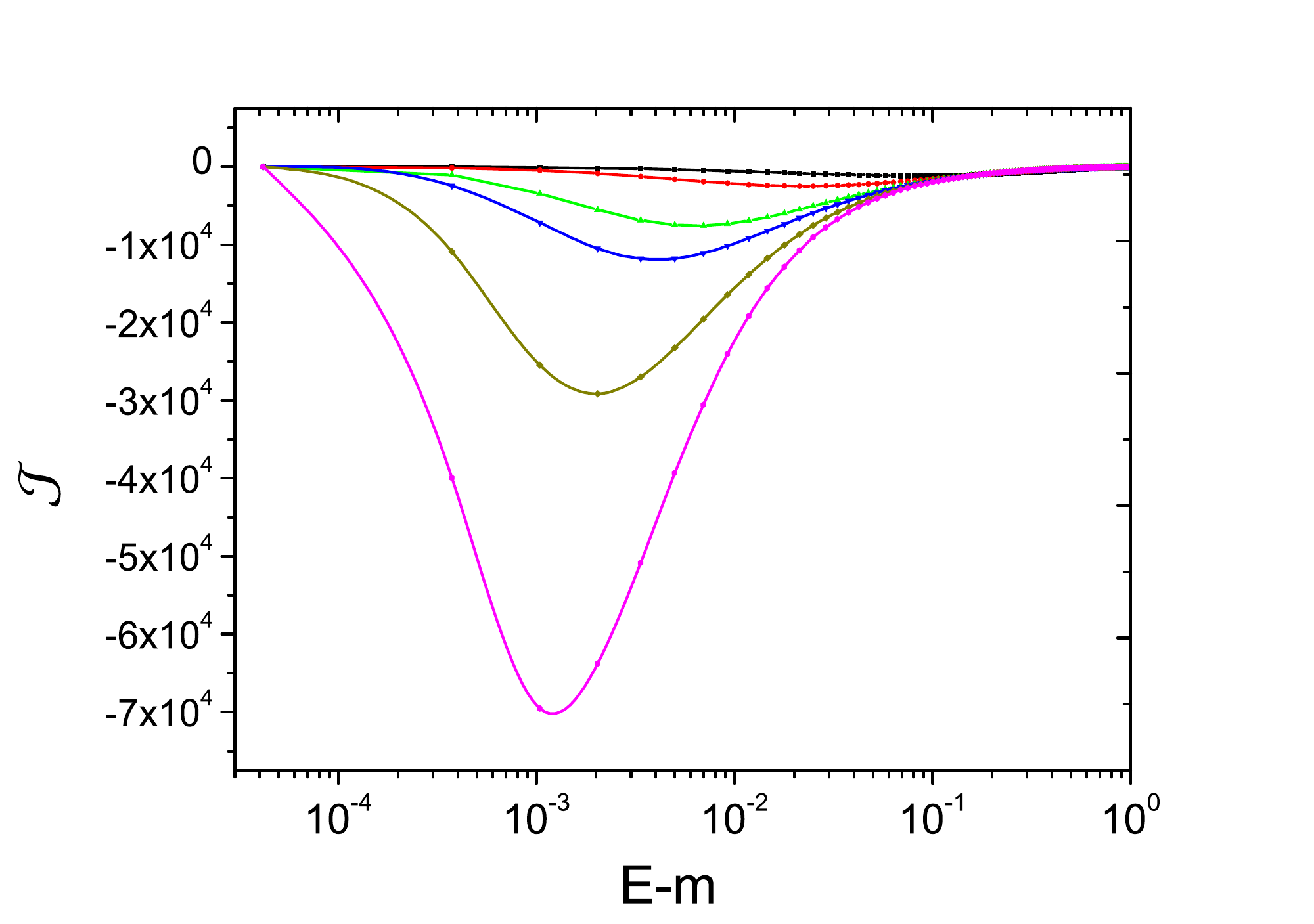}
\caption{{Color online.} Distribution function $f$ (left) and current $[\cal J]$ (right), the curves corresponding, from bottom to top,  to increasing values of $\tau$ : $\tau=$0.04(black), 0.0674(red), 0.0871(green), 0.0927(blue), 0.0961(yellow), 0.0977(pink) with $m=0.3$ and $f_0=1.0$ (overpopulated initial condition). \label{fig:overcurrent}}
\end{center}
\end{figure}

\begin{figure}[!htb]
\begin{center}
\includegraphics[width=6.5 cm]{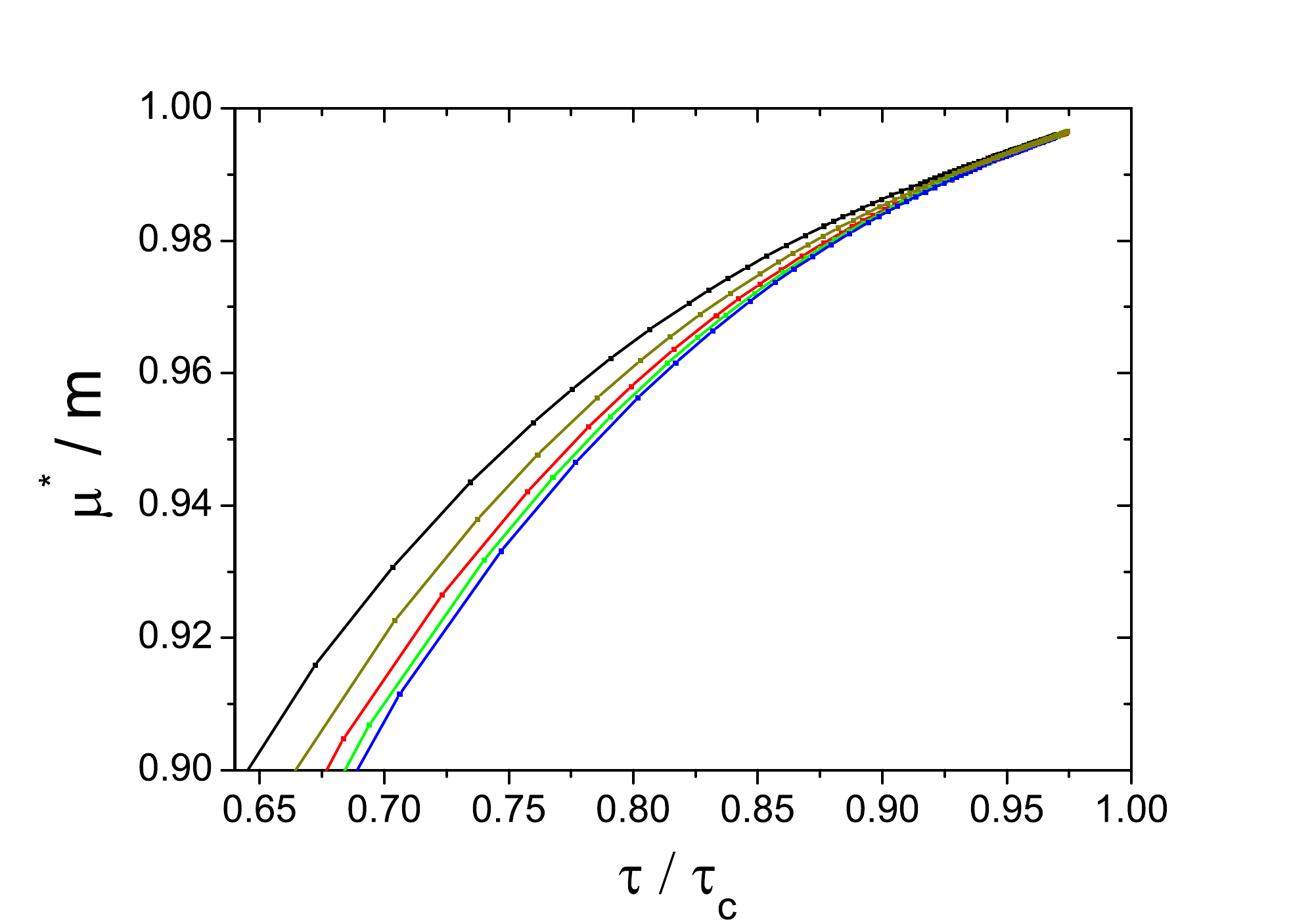}\includegraphics[width=5.5cm]{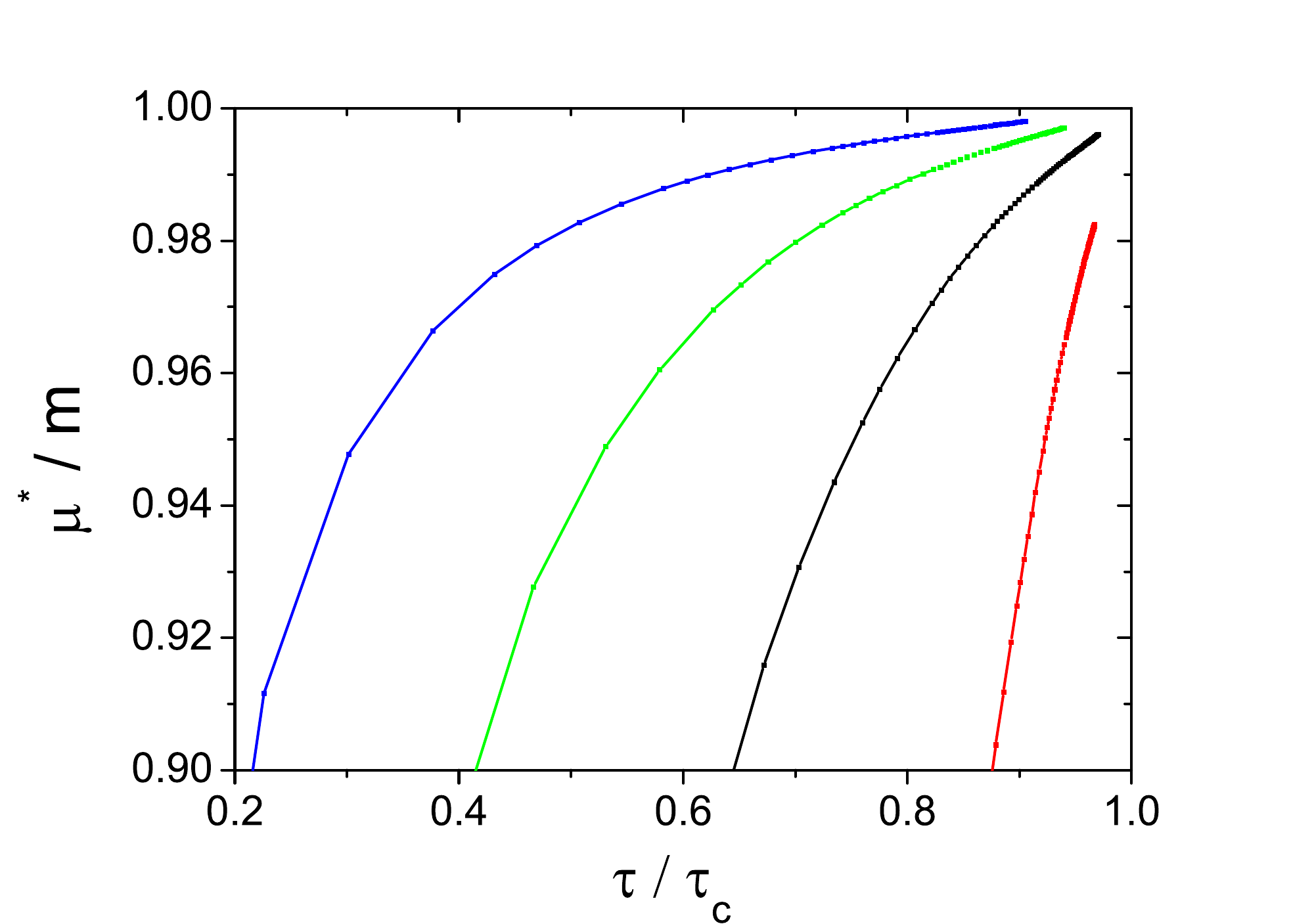}
\caption{Evolution of IR local chemical potential $\mu^*$ toward onset $\mu^*\to m$ for different choices of 
$f_0=$1.0, 1.2, 1.4, 1.6, 1.8(from left to right) with the same mass $m=0.3$. Right: Evolution of IR local chemical potential $\mu^*$ toward onset $\mu^*\to m$ for different choices of 
external mass $m=$0.7, 0.5, 0.3, 0.1(from left to right) with the same $f_0=1$. \label{fig:muonset} }
\end{center}
\end{figure}

Different from the underpopulated case, the evolution of the flow of particles is quite different in the overpopulated case, as shown in Fig.\ref{fig:overcurrent} . The negative infrared flux keeps increasing in a roughly self-similar manner. The local chemical potential $\mu^*$   approaches $\mu_c=m$ at the onset.  We have extracted the $\mu^*$ as a function of time for a variety of mass values as well as initial occupation values: see Fig.\ref{fig:muonset}. In all cases, we have found that the evolution of $\mu^*$ close to the onset point can be fitted with power law $\mu=m- \lambda \, (\tau_c-\tau)^\eta$. Furthermore in all cases, we've found the optimal exponent is about one, i.e. $\eta \simeq 1$. 
 This dynamical behavior in the present massive case is essentially the same as that found for the massless case in \cite{Blaizot:2013lga}. We give  a more complete discussion of this regime in the next section.

\section{Critical scaling analysis}\label{Sec:scaling}

In this section we examine how the scaling behavior that holds near the onset of BEC is affected by the gluon mass. A major modification, with respect to the massless case,  concerns the  dispersion relation  near the onset. When $m\ne 0$, condensation occurs when $m=\mu$, and the dispersion relation becomes non relativistic,  $E_p-\mu=\sqrt{p^2+m^2}-m\approx p^2/(2m)$, which differs from the ultra relativistic dispersion relation of the massless case, $E_p=p$,  where condensation occurs at $\mu=0$. This change in the dispersion relation modifies the singularity near the onset, but, as we shall see, this does not affect in a major way the onset critical behavior.  The foregoing analysis  follows closely that presented in Ref.~\cite{Blaizot:2013lga}.

\subsection{Scaling behavior of the current at small momentum}

As we have seen in solving numerically the Fokker-Planck equation, for generic initial conditions,  the distribution function at small momentum evolves rapidly towards  the approximate equilibrium distribution: 
\begin{eqnarray}\label{fclassical}
f(p) \approx \frac{T^*}{E(p)-\mu^*} \approx \frac{2 mT^*}{ p^2 + 2m\delta\mu} =\frac{2 mT^*}{ p^2 + \Delta^2},
\end{eqnarray}
where we have set $\delta\mu \equiv m-\mu^*$ and $\Delta\equiv\sqrt{2m\delta\mu}$. As the system approaches condensation, $\Delta\to 0$. To analyze the detailed behavior of the properties of the system when $\Delta \to 0$, it is convenient to calculate the time dependence of the number of particles inside  a small sphere of radius $p_0$ centered at the origin. In doing this calculation, we assume that $f(p)$ keeps the form (\ref{fclassical}), with time dependent parameters $T^*$ and $\delta\mu$. In particular, $f(0)$ remains finite as long as $\delta\mu\ne 0$. A simple calculation then yields
\begin{eqnarray}\label{smallp0}
\partial_\tau \int_0^{p_0} dp p^2 f(p) = 2m \Delta (\partial_\tau T^*) h_1 (y) + 2m T^* (\partial_\tau \Delta) h_2(y) ,
\end{eqnarray}
where  $y\equiv p_0/\Delta$, and the two scaling functions are
\begin{eqnarray}
h_1 (y) &=& y - {\rm ArcTan}(y) \\
h_2 (y) &=& \frac{y}{1+y^2} - {\rm ArcTan}(y).
\end{eqnarray}
By using Eq.~(\ref{spehricaleqn}), one can relate the left hand side of Eq.~(\ref{smallp0}) to the current at $p_0$, and obtain
\begin{eqnarray}
-{\mathcal J} (p_0) = \frac{\partial_\tau \int_0^{p_0} dp p^2 f(p)}{p_0^2}=\frac{2m(\partial_\tau T^*)}{\Delta} \frac{h_1(y)}{y^2} + \frac{2m T^* (\partial_\tau \Delta)}{\Delta^2}  \frac{h_2(y)}{y^2}.   \label{eq_J_h1h2}
\end{eqnarray}
This equation provides interesting constraints on the small $p$ behavior of ${\cal J}(p)$.

\begin{figure}[!hbt]
\begin{center}
\includegraphics[width=0.5\textwidth]{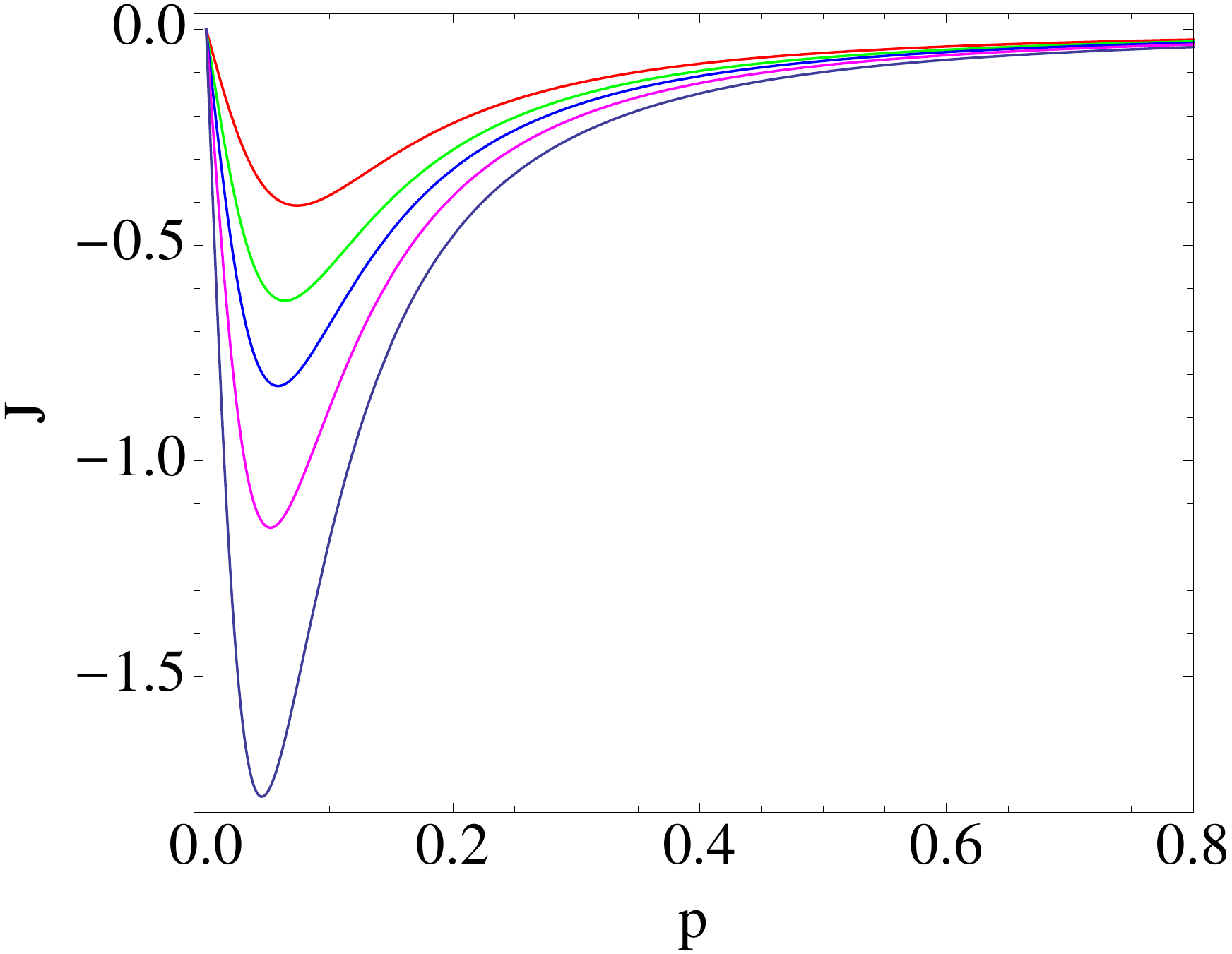}\includegraphics[width=0.6\textwidth]{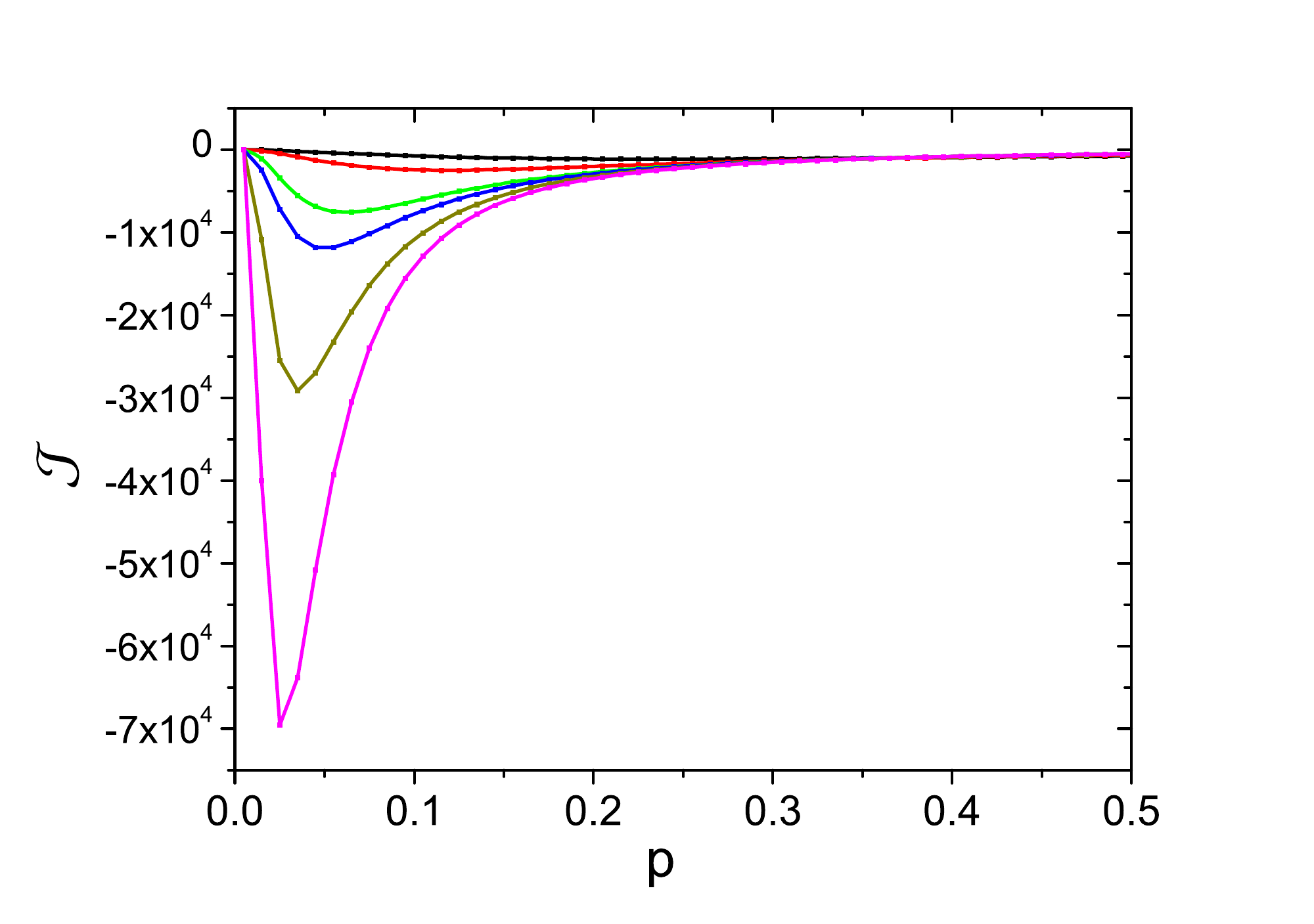}
\caption{The current (in arbitrary units) as a function of momentum, at different time moments (earlier to later time from top to bottom) close to onset. Left panel: calculated according to the formula (\ref{eq_J_h1h2}), assuming a linear relation between $\mu^*-m$ and $\tau_c-\tau$, and neglecting the time variation of $T^*$; right panel: obtained from the numerical solution. The linear behavior at small $p$ followed, as $p$ increases, by a singular behavior in $1/p^\alpha$, with $\alpha>0$ is clearly visible on both figures. } \label{fig:onset_current}
\end{center}
\end{figure}

 In the limit $y\to 0$ or $p_0 \ll  \Delta$, one gets 
\begin{eqnarray}\label{currentp0}
-{\mathcal J} (p_0) \simeq p_0 \times \frac{2m}{3} \partial_\tau \left( \frac{T^*}{\Delta^2} \right) =
p_0 \times \frac{1}{3} \partial_\tau \left( \frac{T^*}{\delta\mu} \right)  = \frac{p_0}{3} \partial_\tau f(0).
\end{eqnarray}
The current is linear in $p_0$, with a slope proportional to $\del_\tau f(0)$, very much like in the massless case \cite{Blaizot:2013lga}. On the other hand, in the limit $y\gg 1$ or $\Delta\ll  p_0 \ll T^*$, one obtains the following leading order result\footnote{The subleading contribution is $\sim 1/p$ and comes from the first term in Eq.~(\ref{eq_J_h1h2}). It becomes the dominant contribution after onset, when $\delta\mu=0$ and therefore (\ref{currentp0b}) vanishes.}
\begin{eqnarray} \label{currentp0b}
-{\mathcal J} (p_0) \to \frac{1}{p_0^2} \times (\pi m) \partial_\tau (-T^* \Delta).
\end{eqnarray}
In this region, the current exhibits a singular behavior in $1/p_0^2$. 
The  value ($\approx \Delta$) of the momentum where the change of regime occurs decreases with time, while the absolute value of the current at the minimum becomes larger and larger, and eventually diverges at the onset. This behavior is illustrated in Fig.~\ref{fig:onset_current}. As was the case in the massless case, the current in the scaling regime is dominated by the second term in Eq.~(\ref{eq_J_h1h2}). One also finds that the dependence of $m-\mu^*$ on $\tau_c-\tau$ is linear. A discussion of the latter point is presented in Appendix~\ref{Sect:smallmomentumJ}. 

\section{Summary}

In this paper, we have extended the analysis carried out in Ref.~\cite{Blaizot:2013lga} to the case where the gluons are given a small mass $m$. The main difference with respect to the massless case studied in \cite{Blaizot:2013lga} is the change of the dispersion relation of the gluon modes, from ultra-relativistic in the massless case to non-relativistic in the massive case. While this  change in   the dispersion relation leads to an enhanced singularity of the distribution function near the onset of Bose condensation, from $1/p$ to $1/p^2$, we find that the critical regime that accompanies the approach to the onset  is qualitatively unchanged. The Debye mass $m_D$ controls the infrared behavior of the collision kernel. It is kept constant, and independent of $m$, although in a more complete theory, both masses are related, and would depend on the temperature. However this is expected to have little impact at the onset, since the onset process is dominated by a critical regime where the actual value of the mass is irrelevant. We have checked in articular that letting the Debye mass adjust with temperature as one approaches BEC does not lead to any significant changes in the  behavior of the system. Finally, the presence of the mass $m$ allows us to define properly the equations that governs the dynamics beyond the onset, that is, in the presence of the condensate. This is discussed in a companion paper \cite{BLM2}.

\section*{Acknowledgements}
The research of JPB is 
supported by the European Research Council under the
Advanced Investigator Grant ERC-AD-267258. That of YJ and JL is supported by the National Science Foundation under Grant No. PHY-1352368. JL is also grateful to the RIKEN BNL Research Center for partial support. 

\begin{appendix}


\section{Derivation of the current ${\cal J}(\p)$}

The component $i$ of the current ${\cal J}(\p_1)$  in Eq.~(\ref{currentJi}) can be written
\beq
{\cal J}^i(\p_1) &=& 18\pi {g^4}\int {\frac{{{\rmd^3}\q}}{{{{(2\pi )}^3}}}} \int {\frac{{{\rmd^3}{\p_2}}}{{{{(2\pi )}^3}}}} \,({h_1}{f^\prime_2} - {h_2}{f^\prime_1})\\\nonumber
&\times&\q^i (\q \cdot {{\boldsymbol v}_2})\,\frac{{{{(1 - {{{\boldsymbol v}}_1} \cdot {{{\boldsymbol v}}_2})}^2}}}{{{{[{{({{{\boldsymbol v}}_1} \cdot \q)}^2} - {\q^2} - {m_D^2}]}^2}}}\,\delta (\q \cdot {{\boldsymbol v}_1} - \q \cdot {{\boldsymbol v}_2}).
\eeq

\subsection{Details of the angular integration}

In order to perform the angular integrals, we choose the following coordinate frames: We define the orientation of $\q$ in a frame where $\p_1$ is along the $\hat{z}$ axis,  and denote  the corresponding angles by ${\theta}$ and ${\phi}$. The orientation of $\p_2$, given $\p_1$ and $\q$, is defined by the  angles $\theta_2,\phi_2$ in a frame with $\q$ along $\hat{z}_2$ and  $\q,\p_1$ spanning the $\hat{x}_2$-$\hat{z}_2$ plane.
We have therefore  $\q\cdot {\boldsymbol v}_1=q v_1 \cos{{\theta}}$, $\q\cdot {\boldsymbol v}_2=q v_2 \cos\theta_2$,
and ${\boldsymbol v}_1 \cdot {\boldsymbol v}_2 = v_1 v_2 (\cos{{\theta}}\cos\theta_2 + \sin{{\theta}}\sin\theta_2 \cos\phi_2 )$. 

By integrating over $\phi_2$ one gets
\begin{eqnarray}
\nonumber
&\int_0^{2\pi } {d{\phi _2}} {(1 - {{{\boldsymbol v}}_1} \cdot {{{\boldsymbol v}}_2})^2} = \int_0^{2\pi } {d{\phi _2}} {[1 - {v_1}{v_2}(\cos \theta \cos {\theta _2} + \sin \theta \sin {\theta _2}\cos {\phi _2})]^2}\\
 &= \pi [2{(1 - {v_1}{v_2}\cos \theta \cos {\theta _2})^2} + v_1^2v_2^2(1 - {\cos ^2}\theta )(1 - {\cos ^2}{\theta _2})]
\end{eqnarray}
Furthermore we rewrite the delta function as
\begin{eqnarray}
q\delta (\q \cdot {{{\boldsymbol v}}_1} - \q \cdot {{{\boldsymbol v}}_2}) = q\delta (q({v_1}\cos \theta  - {v_2}\cos {\theta _2})) = \delta ({v_1}\cos \theta  - {v_2}\cos {\theta _2})\\\nonumber
 = \theta ({v_1} - {v_2})\delta ({v_1}\cos \theta  - {v_2}\cos {\theta _2}) + \theta ({v_2} - {v_1})\delta ({v_2}\cos {\theta _2} - {v_1}\cos \theta )
\end{eqnarray}
and change  variables, setting ${x_1} = {v_1}\cos \theta$, ${x_2} = {v_2}\cos {\theta _2}$. One can then complete the angular integration. Note that, by symmetry, the current is aligned on
the direction of ${\p}_1$, i.e.,  $\vec{\cal J}(\p_1)=\hat\p_1{\cal J}(p_1)$, with ${\cal J}(p_1)$ a function of $p_1=|\p_1|$ only, and $\hat\p_1=\p_1/p_1$..  We get
\beq\label{Jp1}
{\cal J}(p_1)=\frac{18\alpha_s^2}{\pi}\int \rmd p_2 \,p_2^2 \,({h_1}{{f^\prime}_2} - {h_2}{{f}^\prime_1})\int\frac{\rmd q}{q}\frac{Z(v_1,v_2,c_q)}{v_1},
\eeq
with
\beq\label{v1v2Z}
v_1v_2\,Z(v_1,v_2,c_q)&=&\frac{1}{2}\left[\int_{ - {v_2}}^{{v_2}} {\rmd{x_2}}\, x_2^2\,\frac{{2{{(1 - x_2^2)}^2} + (v_1^2 - x_2^2)(v_2^2 - x_2^2)}}{{{{(x_2^2 - c_q)}^2}}}\,\theta ({v_1} - {v_2}) \right.\nn
&+&\left. \int_{ - {v_1}}^{{v_1}} {\rmd{x_1}} x_1^2\,\frac{{2{{(1 - x_1^2)}^2} + (v_1^2 - x_1^2)(v_2^2 - x_1^2)}}{{{{(x_1^2 - c_q)}^2}}}\,\theta ({v_2} - {v_1})\right]\nn
&=&  v\Big[{v^2} - 6(1 - c_q)+(1 - v_1^2) + (1 - v_2^2) \nn
& +&  \left.\frac{{{{(1 - c_q)}^2}}}{{c_q - {v^2}}} + \frac{{(c_q - v_1^2)(c_q - v_2^2)}}{{2(c_q - {v^2})}}\right]\nn
&+& \left[ - \frac{{{{(1 - c_q)}^2}}}{{c_q}} - \frac{{(c_q - v_1^2)(c_q - v_2^2)}}{{2c_q}} - (1 - v_1^2) - (1 - v_2^2) \right.\nn
&+& 6(1 - c_q)\Big]\sqrt {c_q}\,{\rm ArcTanh}\left(\frac{v}{{\sqrt {c_q} }}\right),
\eeq
where $c_q = 1 + {m_D^2}/{q^2}$, $v = \min ({v_1},{v_2})$, and we have used the following integral formula to get the final results
\begin{eqnarray}
& &\int {\rmd x} \,{x^2}\left[3C + \frac{B}{{{c_0} - {x^2}}} + \frac{A}{{{{({c_0} - {x^2})}^2}}}\right]\\\nonumber
&  &= C{x^3} - Bx - \frac{{Ax}}{{2({x^2} - {c_0})}} + \left(B\sqrt {{c_0}}  - \frac{A}{{2\sqrt {{c_0}} }}\right){\rm ArcTanh}\left(\frac{x}{{\sqrt {{c_0}} }}\right),
\end{eqnarray}
with $A = 2{({c_0} - 1)^2} + ({c_0} - v_1^2)({c_0} - v_2^2)$,  $B = 4 + v_1^2 + v_2^2 - 6{c_0}$, $C = 1$.

\subsection{Limiting cases}

There are a few limiting cases where the expression above simplifies. For instance, in the limit of vanishing screening mass $m_D\to 0$, $c_q=1$ and
\begin{eqnarray}
{v_1}{v_2}Z& =& v\left[{v^2} + (1 - v_1^2) + (1 - v_2^2)  - \frac{{(1 - v_1^2)(1 - v_2^2)}}{{2(1 - {v^2})}} \right]\nn
 &+&  \left[  \frac{{(1 - v_1^2)(1 - v_2^2)}}{{2}} - (1 - v_1^2) - (1 - v_2^2) \right]{\rm ArcTanh}(v).
\end{eqnarray}
In this limit, the remaining $q$-integration in the expression (\ref{Jp1}) of the current  becomes simply the usual Coulomb logarithm, $\int dq/q \to {\cal L}$. 

In the limit where all effective masses vanish, i.e. $m\to 0$ and $m_D\to 0$,  $c_q$ and all the velocities become unity,  $v_1 v_2 Z \to 1$ and we recover the expression of the massless case studied in Ref.~\cite{Blaizot:2013lga}.  

In the limit where the colliding gluons become massless, i.e.,  $m \to 0$, but the Debye mass stays finite, all the velocities $v_1,v_2,v$ approach unity, and we get
\begin{eqnarray}
{v_1}{v_2}Z& = &  \left[1 - \frac{15}{2}(1-{c_q})\right] \nn
&&+ \sqrt {{c_q}}(1-c_q) \left[ - \frac{{{{3(1 - {c_q})}}}}{{{2c_q}}} + 6\right]{\rm ArcTanh}\left(\frac{1}{{\sqrt {{c_q}} }}\right).
\end{eqnarray}
This expression is finite when $q\to 0$. To see that,  recall that when $q\to 0$, $c_q \sim m_D^2/q^2 \to \infty$. As simple calculation then yields 
\begin{eqnarray}
{v_1}{v_2}Z\approx  A\,c^{-2}_q+{\cal O}(c_q^{-3}),
\end{eqnarray}
where
\begin{eqnarray}
A= \frac{3}{7} v^7-\frac{v^5}{5}(4+v^2_1+v^2_2)+\frac{v^3}{3} (2+v^2_1v^2_2)].
\end{eqnarray}
This result confirms that in the presence of a non-vanishing screening mass $m_D$,  the function $Z$ is regular as $q\to 0$, and the $q$-integration in the expression (\ref{Jp1}) becomes infrared finite, as expected.

\subsection{Detailed derivation of the $q$-integration}

In the general case, in order to perform the $q$-integration in Eq.~(\ref{Jp1}), we rewrite $v_1v_2Z$  in Eq.~(\ref{v1v2Z}) as follows:
\begin{eqnarray}
v_1v_2Z(c_q,v_1,v_2)&=&\frac{E_1}{c_q-v^2}+E_2(c_q-v^2)+E_3+\sqrt{c_q}{\rm ArcTanh}\left(\frac{v}{\sqrt{c_q}}\right)\nonumber\\
&&\times[E_4(1-c_q)+E_5\frac{c_q-1}{c_q}+E_6],
\end{eqnarray}
where
\begin{eqnarray}
E_1&=&\frac{v}{4}[2v^2(3v^2-v_1^2-v_2^2-4)+2v_1^2v_2^2+4],\qquad 
E_2=\frac{15v}{2}\nonumber\\
E_3&=&\frac{v}{2}(20v^2-3v_1^2-3v_2^2-12),\qquad
E_4=\frac{15}{2}\nonumber\\
E_5&=&\frac{1}{4}[-2v_1^2v_2^2-4],\qquad 
E_6=\frac{1}{4}[-2v_1^2v_2^2+6v_1^2+6v_2^2-10].
\end{eqnarray}
Then, we change integration variable from $q$ to $c_q$ and obtain 
\begin{eqnarray}
&&\int\frac{dq}{q}v_1v_2Z(q,v_1,v_2) = \int\frac{dc_q}{2(1-c_q)}v_1v_2Z(c_q,v_1,v_2) \nonumber\\
&&=E_1G_1(v,c_q)+E_1G_1(v,c_q)+E_2G_2(v,c_q)+E_3G_3(v,c_q)\nonumber\\
&&+E_4G_4(v,c_q)+E_5G_5(v,c_q)+E_6G_6(v,c_q).
\end{eqnarray}
The results of these integrals are listed below:
\begin{eqnarray}
&&G_1(v,c_q)=\int dc_q\frac{1}{2(1-c_q)(c_q-v^2)}=-\frac{\ln(c_q-v^2)-\ln(c_q-1)}{2(v^2-1)}\nn
&&G_2(v,c_q)=\int dc_q\frac{c_q-v^2}{2(1-c_q)}=-\frac{1}{2}[c_q+(1-v^2)\ln(c_q-1)]\nn
&&G_3(v,c_q)=\int dc_q\frac{1}{2(1-c_q)}=-\frac{1}{2}\ln(c_q-1)\nn
&&G_4(v,c_q)=\int dc_q\frac{1}{2}\sqrt{c_q}{\rm ArcTanh}\left(\frac{v}{\sqrt{c_q}}\right)=\frac{1}{6}[2c_q^{3/2}{\rm ArcTanh}\left(\frac{v}{\sqrt{c_q}}\right)+v^3\ln(c_q-v^2)+c_qv]\nn
&&G_5(v,c_q)=\int dc_q\frac{1}{2c_q}{\rm ArcTanh}\left(\frac{v}{\sqrt{c_q}}\right)=\frac{1}{2}\left[v \ln(c_q-v^2)+2\sqrt{c_q}{\rm ArcTanh}\left(\frac{v}{\sqrt{c_q}}\right)\right]\nn
\eeq
\beq
&&G_6(v,c_q)=\int dc_q\frac{1}{2(1-c_q)}{\rm ArcTanh}\left(\frac{v}{\sqrt{c_q}}\right)\nn
&&=v\{-\frac{\sqrt{c_q}}{v}{\rm ArcTanh}\left(\frac{v}{\sqrt{c_q}}\right)+\ln\left(\frac{v}{\sqrt{c_q}}\right)-\frac{1}{2}\ln\left(1-\frac{v^2}{c_q}\right)\nn
&&-\frac{1}{4v}\left[2\ln(\frac{\sqrt{c_q}-1}{\sqrt{c_q}+1}){\rm ArcTanh}\left(\frac{v}{\sqrt{c_q}}\right)+{\rm Li}_2\left(\frac{1-v/\sqrt{c_q}}{1+v}\right)+{\rm Li}_2\left(-\frac{v+v\sqrt{c_q}}{1-v}\right)\right.\nn
&&+{\rm Li}_2(-\frac{v-v\sqrt{c_q}}{1-v})+{\rm Li}_2(\frac{1+v/\sqrt{c_q}}{1+v})\nn
&&+\ln(v-v/\sqrt{c_q})\ln(1-v/\sqrt{c_q})-\ln(v^2-v^2/c_q)\ln(1-v)\nn
&&\left.-\ln(1-v^2/c_q)\ln(1+v)+\ln(v+v/\sqrt{c_q})\ln(1+v/\sqrt{c_q})\right]\}.
\end{eqnarray}
We define, with $\Lambda$ an ultraviolet cutoff\footnote{One expects $\Lambda$ to be typically of the order of the temperature. The numerical calculations have been performed  with a somewhat larger value, $\Lambda=4 Q_s$.}: 
\begin{eqnarray}
\int_0^\Lambda\frac{dq}{q}v_1v_2Z(c_q,v_1,v_2)=L(c_\Lambda,v_1,v_2)-L(\infty,v_1,v_2),
\end{eqnarray}
where the first argument of $L$ denotes the value of $c_q$ (when $q=\Lambda$ or $q=0$, respectively). 
As we have already shown, there is no singularity at $q\to 0$ i.e. $c_q \to \infty$ when $m_D$ is finite,  and we have the following finite result 
\begin{eqnarray}
&&L(\infty,v_1,v_2)=\frac{1}{24}{\bigg \{} 20v^3-12v(v_1^2v_2^2+2)]\nn
&&-3\, \Delta \, \left[{\rm Li}_2\left(\frac{v}{v-1}\right)+{\rm Li}_2\left(\frac{1}{v+1}\right)+2v-(2v+\ln(1-v))\ln(v)\right] {\bigg \}}, \nn
\end{eqnarray}
where $\Delta=-2 v_1^2 v_2^2 + 6(v_1^2 + v_2^2 ) - 10$. Note in the massless limit for external gluons, i.e. $m\to 0$, one has $\Delta \to 0$ and the expression  above greatly simplifies: 
\begin{eqnarray}
L(c_\Lambda)-L(\infty) = -\frac{1}{2}\left[5c_\Lambda+(3-5c_\Lambda)\sqrt{c_\Lambda}{\rm ArcCoth}(\sqrt{c_\Lambda})-\frac{4}{3}\right]. 
\end{eqnarray}

 We write the final result for the current in the form
\begin{eqnarray}
{\cal J}(\p_1)=\frac{18\alpha_s^2}{\pi}\int dp_2 p^2_2 \, {\cal Z}(v_1,v_2) \, \left[ h_1f^\prime_2-h_2f^\prime_1 \right] 
\end{eqnarray}
with 
\beq
{\cal Z}(v_1,v_2) &=&\int\frac{\rmd q}{q} \frac{Z(v_1,v_2,c_q)}{v_1}\nn
&=& (v^2_1v_2)^{-1}\left[L(c_\Lambda,v_1,v_2)-L(\infty,v_1,v_2)\right].
\eeq

 \subsection{The small momentum regime}\label{Sect:smallmomentumJ}

We now analyze the small momentum form of the current,  ${\cal J}(p_1\to 0)$.  The leading order expression  of $\mathcal{Z}$ in the limit  $v_1 \to 0 $ can be obtained after a lengthy but straightforward calculation. It reads
\begin{eqnarray}\label{Zsmallv}
\mathcal{Z}(v_1\to 0, v_2) \approx {\cal L} \,   \frac{v^3}{v_1^2 v_2}  \qquad
\end{eqnarray}
where $v = \min ({v_1},{v_2})$, and  $\cal L$ is a positive constant,
\beq
{\cal L}=-\frac{1}{3}\left[\log (c_\Lambda-1)-\log( c_\Lambda )+\frac{1}{c_\Lambda}\right]   ,\qquad c_\Lambda=1+\frac{m_D^2}{\Lambda^2}.
\eeq
The small momentum current can therefore be written as\footnote{It is not difficult to show that, with this approximate expression for the current, particle number as well as energy are conserved.}
 \begin{eqnarray}
-{\cal J}(p_1) = 36 \pi \alpha_s^2 {\cal L}
  \int_{\p_2}\,  \frac{v^3}{v_1^2 v_2}  \, \left[ h_2{f}^\prime_1 - h_1{f}^\prime_2 \right] .
  \end{eqnarray}
 Since the numerical factor in front of the integral is to be  absorbed in the redefinition of the time scale (see Eq.~(\ref{timescale})), we rewrite ${\cal J}(p_1)$ as
 \beq\label{Jp1tilde}
 -{\cal J}(p_1) =I_a(p_1) {f}^\prime_1 + I_b(p_1)  h_1,
 \eeq
 where
 \beq\label{Iap1Ibp1}
I_a(p_1)= \int_{\p_2} h_2 \frac{v^3}{v_1^2v_2} ,\qquad   I_b(p_1)=- \int_{\p_2} f^\prime_2 \frac{v^3}{v_1^2v_2}
 \eeq
 These integrals reduce to the integrals $I_a$ and $I_b$ of Eq.~(\ref{IaIb}) in the massless limit (in this limit, the dependence on $p_1$ disappears and ${\cal L}$ becomes the usual Coulomb logarithm).

We argue in the main text, and we have verified through numerical calculations,  that at small momentum and near the onset for BEC, the distribution function $f(p)$ can be well approximated by a Bose equilibrium distribution function
We therefore write the distribution in the  momentum range of interest ($p\ll\Delta= \sqrt{2m\delta\mu}$) as
\begin{eqnarray}\label{distribg}
f(p)=f^*(p)+\delta f(p),
\end{eqnarray}
where $f^*(p)$ is an equilibrium Bose distribution with temperature $T^*$ and chemical potential $\mu^*$, and $\delta f(p)$ represents a small deviation from this equilibrium distribution. We assume that $\delta f(p)$ is a regular function of $p$ at $p=0$. 





In order to calculate the integrals $ I_{a,b}$ we need to pay attention to the fact that $v = \min ({v_1},{v_2})$, and divide the $p_2$ integration range appropriately, making explicit the dependence on $p_1$. We obtain
\begin{eqnarray}\label{Iap1}
 2\pi^2\,I_a(p_1)&&=\int_0^{p_1}dp_2 p_2^2 \frac{v_2^2}{v_1^2}h_2 + \int_{p_1}^{\infty}dp_2 p_2^2\frac{v_1}{v_2}h_2\nonumber\\
&&=\int_0^{p_1}dp_2 p_2^2 \left( \frac{v_2^2}{v_1^2}-\frac{v_1}{v_2}\right)h_2 + \int_{0}^{\infty}dp_2 p_2^2\frac{v_1}{v_2}h_2\nn
&&=\int_0^{p_1}dp_2 p_2^2 \left( \frac{p_2^2}{p_1^2}-\frac{p_1}{p_2}\right)h_2 + \frac{p_1}{m}\int_{0}^{\infty}dp_2 p_2 E_2 h_2.
\end{eqnarray}
In the first integral, we could set $v_1\approx p_1/m$ and $v_2=p_2/m$, while in the second integral we need to keep the exact expression $v_2=p_2/E_2$.  Since, when $p_1\ll \sqrt{2m\delta\mu}$,  $h_2$ in the first integral is nearly a constant, this  first integral is of order $p_1^3$, and is therefore negligible compared to the second one $\propto p_1/m$. We are then left with the integral given in Eq.~(\ref{integralstilde}). 
Proceeding in the same way for $-  I_b$, one obtains an identical expression to that of $ I_a$ with  $f^\prime_2$  substituted to $h_2$. We shall denote by $ I_a^*$ and $ I_b^*$, the integrals $ I_a$ and $ I_b$ calculated with the distribution $f^*$, and write accordingly $ I_{a,b}= I_{a,b}^*+\delta  I_{a,b}$. We may then expand the current (\ref{Jp1tilde}) as follows
\beq
 -{\cal J}(p_1) =\delta  I_a  (f_1^*)^\prime +\delta  I_b h_1^*+ I_a^* (\delta f_1)^\prime+ I_b^* \delta h_1,
 \eeq
 where $\delta h_1\approx \delta f_1 (1+2 f_1^*)$, and  we have used the fact that $ I_a^*=T^*  I_b^*$ which follows immediately from the relation $(f^*)^\prime=-(1/T^*) h^*$. By using again the same identity for the first two terms of the equation above, we can rewrite the current as 
 \beq\label{currentJp1small}
 {\cal J}(p_1)=\frac{1}{T^*} (\delta  I_a-T^*\delta I_b) h_1^*- I_a^* (\delta f_1)^\prime- I_b^* \delta h_1.
 \eeq
 At this point, we note that the last two terms in the expression above can be neglected as $p_1\to 0$. Indeed, $\delta f(p_1)$ is regular as $p_1\to 0$, while $h_1^*\sim (f_1^*)^2$ diverges ($\delta h_1\sim \delta f_1 f_1^*$ is subleading). 
A simple calculation yields 
\beq
\delta\tilde I_a-\delta\tilde I_b T^*\approx\frac{p_1}{m}\int_0^{\infty}dp_2 p_2 E_2\left[\delta f_2(1+2f_2^*)+T^*(\delta f_2)^\prime\right]. 
\eeq
Thus, after dropping the last two terms in Eq.~(\ref{currentJp1small}) one can rewrite ${\cal J}$ as follows
\begin{eqnarray}\label{smallp1J}
\mathcal{J}(p_1)=\frac{1}{3}\gamma(\tau) p_1 f^2(p_1), 
\end{eqnarray}
where $\gamma(\tau)$ is a priori a regular function of time which can be expanded  around the onset time $\tau=\tau_c$, $\gamma(\tau)\simeq\gamma_c+\alpha(\tau_c-\tau)$.
\begin{figure}[!hbt]
\includegraphics[width=0.5\textwidth]{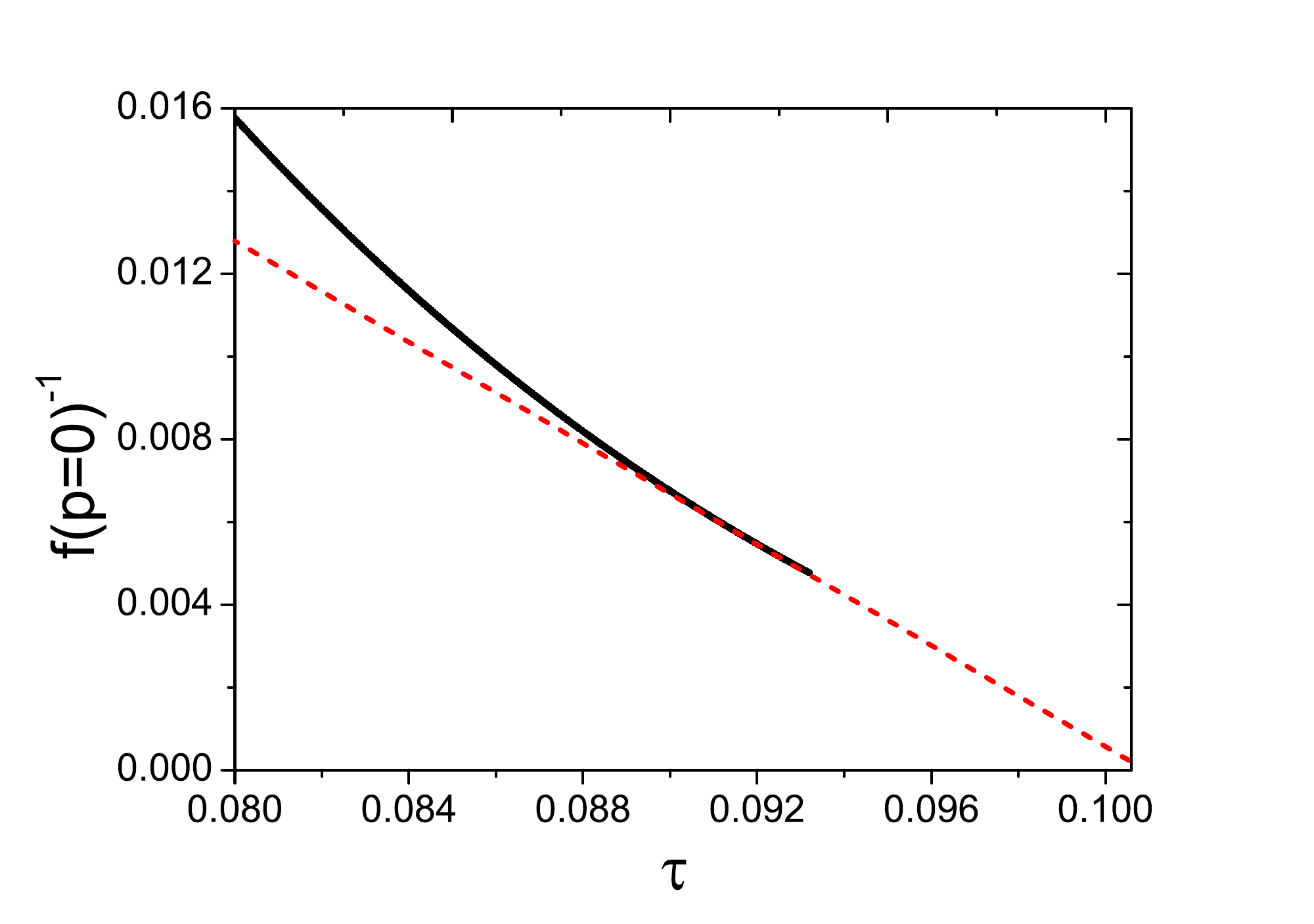}
\includegraphics[width=0.5\textwidth]{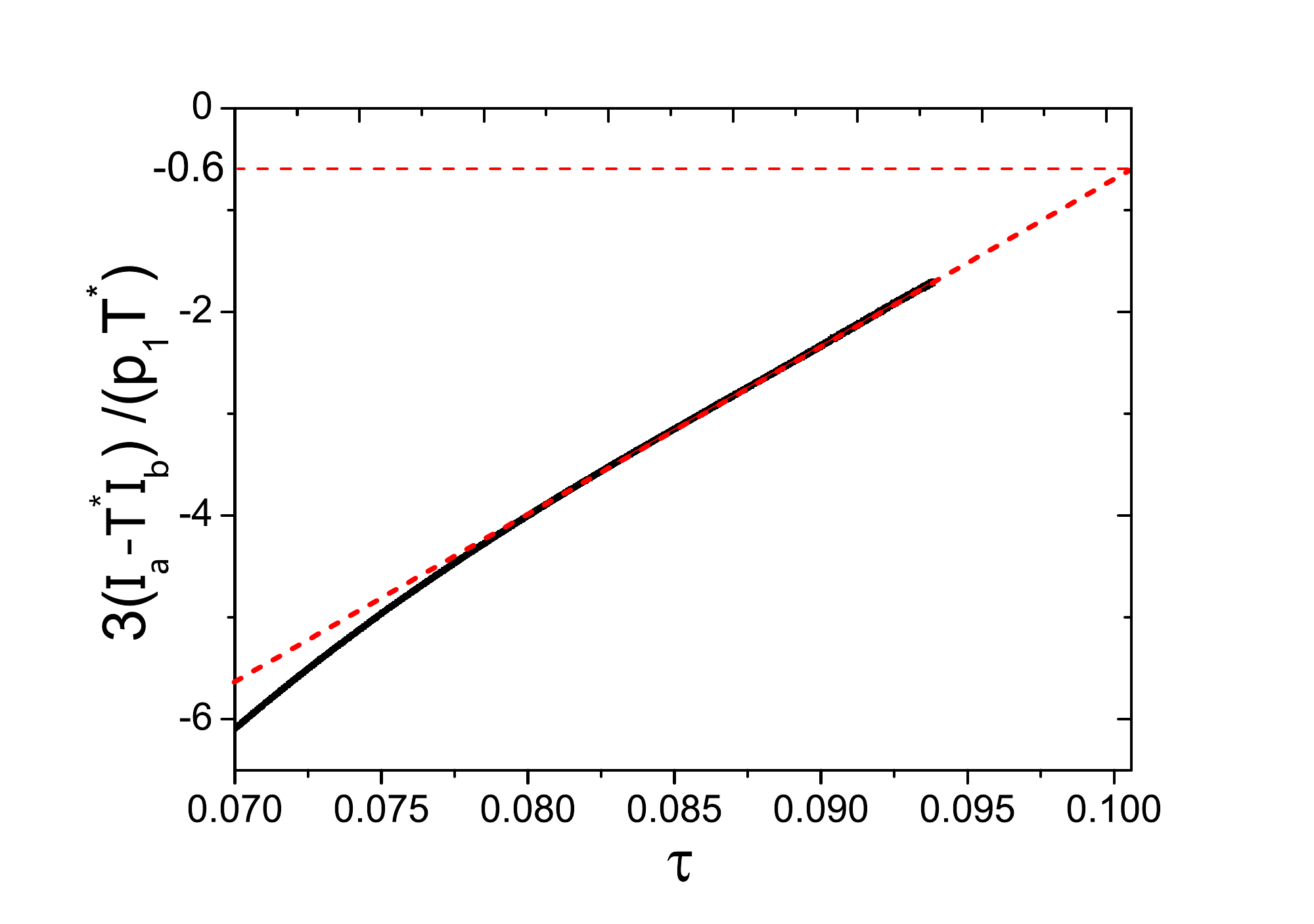}
\caption{$f^{-1}_{p_1}$(left) and $\gamma$(right) as function of time with $m=$0.3.
Red lines are linear fitting results. \label{scalingtest}}
\end{figure}
 By inserting this expression in Eq.~(\ref{currentp0}) one gets the following equation for $f(0)\simeq f(p_1)$
\begin{eqnarray}
\partial_\tau f^{-1}_{p_1}=\gamma(\tau), 
\end{eqnarray}
whose solution reads 
\begin{eqnarray}
f^{-1}_{p_1}(\tau)\simeq-\gamma_c(\tau_c-\tau)-\frac{1}{2}\alpha(\tau_c-\tau)^2
\end{eqnarray}
where we have used the expansion of $\gamma(\tau)$ and  the condition $f^{-1}_{p_1}(\tau_c)\simeq 0$.
We can check numerically that this is the correct behavior. In order to do so, 
we first fit the calculated function $f^{-1}_{p_1}(\tau)$ near $\tau_c$ with a linear form, $\gamma_c(\tau_c-\tau)$, and  get the onset time $\tau_c\simeq 0.1006$ as well as $\gamma_c\simeq -0.6$. Then we compare these values to those obtained from the fit of the quantity $3(I_a-I_b T^*)/(p_1 T^*)$ which exhibits a linear time dependence near $\tau_c$, as expected. The two values agree perfectly, as can be seen in Fig~\ref{scalingtest}. 

It is interesting to compare the present analysis to the analogous one presented in  Ref.~\cite{Blaizot:2013lga}. In this case, $p_1/m\to 1$, and the combination $I_a-T^* I_b$ vanishes.  Both in the massive and the massless case, one finds that $\mu^*-m$ vanishes linearly with $\tau_c-\tau$, but while this result follows from a simple argument in the massive case, in the massless case this could only be determined numerically.

\section{Appendix B: Details of Numerics} \label{Sect:numerics}

In this Appendix, we give some details on how we solve the Fokker-Planck equation. An efficient strategy is to solve its ``integrated version''. Namely, instead of solving the differential equation for the distribution function directly, we evaluate the total number of particles in a thin shell $\int^{p+\Delta p/2}_{p-\Delta p/2}dp p^2 f(p)\approx p^2\Delta p f(p)$ and examine its time evolution by integrating the transport equation over this moment window: 
\begin{eqnarray}
p^2\Delta p \partial_\tau f(p)=(p^2\mathcal{J})^{p+\Delta p/2}_{p-\Delta p/2}
=\mathcal{F}(p+\Delta p/2)-\mathcal{F}(p-\Delta p/2)
\end{eqnarray}
Note on the right-hand side the kernel (taking the form of a full derivative $\sim \bigtriangledown \cdot \mathcal{J}$) will be integrated to give the difference of the flux $\mathcal{F} =p^2 \mathcal{J}$ on the two surfaces of this shell at $p\pm \Delta p/2$ respectively. 

Numerically we discretize the distribution on an equally spaced momentum grid
$p_i = (i - 1/2)\Delta p$, $i = 1, ..., 400$, where $\Delta p = 0.01$. So the flux is on the grid
$p_i = i \Delta p$, $i = 0, 1, ..., 400$. It is easy to see that $\mathcal{F}(p = 0) = 0$. We further set the flux to vanish at our momentum grid's UV cutoff, i.e. 
$\mathcal {F}(p = \Lambda = 4Q_s) = 0$, as the boundary condition that enforces exact particle number conservation. To solve this equation we use the
implicit Gaussian scheme (which is a standard algorithm for this type of equation) as
\begin{eqnarray}
f_{\tau+\delta\tau}(p_i)-f_{\tau}(p_i)=\frac{\delta\tau}{p^2\Delta p}
(\mathcal{F}_{p_i+\Delta p/2}[f_{\tau+\delta\tau}]-\mathcal{F}_{p_i-\Delta p/2}[f_{\tau+\delta\tau}])
\end{eqnarray} 
where $\mathcal{F}_p[f_{\tau+\delta\tau}]$ is the flux at momentum $p$ evaluated  with the distribution at $\tau +\delta\tau$. The implicit scheme then involves numerically solving the above set of equations to extract the distribution at time $\tau+\delta\tau$. The advantage of this implicit scheme, as is well know, is its robust numerical stability as compared with e.g. the explicit scheme of directly evolving the equation in time. During
the whole evolution we have implemented an automated adjustment of the time step $\delta\tau$ to guarantee that the distribution function at the lowest grid point (which has the largest occupation $f(p)$)
 $p = \Delta p/2$ varies less than 5\% at each time step forward. This allows rather accurate handling of the very infrared part of the evolution which is important for understanding the critical behavior in the onset of condensation. In the entire calculation the particle number conservation is exact while the energy conservation is maintained at the order of $10^{-3}$ variation or less.
 
\end{appendix}

\end{document}